\documentclass[manuscript]{aastex}

\usepackage{graphicx}
\usepackage{natbib}
\usepackage{latexsym,amssymb}
\usepackage{amsmath}
\usepackage{cases}
\usepackage{mathrsfs}
\usepackage{subfigure}

\newcommand{\dgr}{$^\circ$}

\newcommand{\drm}{\mathrm{d}}                      
\begin{document}

\title{The Gravitational Potential Near the Sun From SEGUE K-dwarf Kinematics}

\author{LAN ZHANG \altaffilmark{1,2,3}, HANS-WALTER RIX \altaffilmark{3}, GLENN VAN DE VEN \altaffilmark{3}, JO BOVY\altaffilmark{4,5}, CHAO LIU \altaffilmark{1},  and GANG ZHAO\altaffilmark{1}}

\altaffiltext{1}{Key Lab of Optical Astronomy, National Astronomical Observatories, CAS, 20A Datun Road, Chaoyang District, 100012 Beijing, China}
\altaffiltext{2}{Graduate University of the Chinese Academy of Sciences, 19A Yuquan Road, Shijingshan District, 100049 Beijing, China}
\altaffiltext{3}{Max-Planck-Institute for Astronomy, K\"{o}nigstuhl 17, D-69117 Heidelberg, Germany}
\altaffiltext{4}{Institute for Advanced Study, Einstein Drive, Princeton, NJ 08540, USA}
\altaffiltext{5}{Hubble fellow}

\begin{abstract}
To constrain the Galactic gravitational potential near the Sun ($\sim$1.5~kpc), we derive and model the spatial and velocity distribution for a sample of 9000 K-dwarfs with spectra from SDSS/SEGUE, which yield radial velocities and abundances ([Fe/H] \& [$\alpha$/Fe]). We first derive the spatial density distribution for three abundance-selected sub-populations of stars accounting for the survey's selection function. The vertical profile of these sub-populations are simple exponentials and their vertical dispersion profile is nearly isothermal.
To model these data, we apply the `vertical' Jeans Equation, which relates the observable tracer number density and vertical velocity dispersion to the gravitational potential or vertical force. We explore a number of functional forms for the vertical force law, and fit the dispersion and density profiles of all abundance selected sub-populations simultaneously in the same potential, and explore all parameter co-variances using MCMC. Our fits constrain a disk {\it mass} scale height $\lesssim$ 300~pc and the total surface mass density to be $67 \pm 6~M_{\odot}\,{\rm pc^{-2}}$ at $|z| = 1.0$~kpc of which the contribution from all stars is $42 \pm 5~~M_{\odot}\,{\rm pc^{-2}}$ (presuming a contribution from cold gas of $13~M_{\odot}\,{\rm pc^{-2}}$).
We find significant constraints on the local dark matter density of $0.0065\pm0.0023 ~M_{\odot}\,{\rm pc^{-3}}$ ($0.25\pm0.09~{\rm GeV\,cm^{-3}}~$). Together with recent experiments this firms up the best estimate of $0.0075\pm0.0021 ~M_{\odot}\,{\rm pc^{-3}}$ ($0.28\pm0.08~{\rm GeV\,cm^{-3}}~$), consistent with global fits of approximately round dark matter halos to kinematic data in the outskirts of the Galaxy.

\end{abstract}


\section{Introduction}
\label{sec:intro}
Since the works of \citet{oort32,oort60}, determining the Galactic gravitational potential from the distribution and kinematics of stars has been one of the most instructive problem in Galactic disk study. Comparing it with the mass density distribution of visible material (stars and gas), one can derive the distribution of dark matter in the Galactic disk. Comparisons of such local dark matter estimations with rotation curve \citep{webe10} constrain the shape of the dark matter distribution. 

\citet[][hereafter KG89a,b,c]{kui89ii,kui89iii,kui89i} pioneered a practical approach to constrain the Galactic potential near the Sun. They used K-dwarfs to estimate the surface mass density of the total gravitating mass and the mass density of dark matter by solving the `vertical' Jeans' equation (i.e. in the 1D equation in the $\hat{\bf z}$-direction), which correlates the space density and velocity of tracer stars. They found the identified (i.e. stellar and gaseous) surface density and the total surface mass density of all gravitating matter within $|z| \leq 1.1$~kpc from the Galactic plane near the Sun to be $48 \pm 8~M_{\odot}\,{\rm pc^{-2}}$ and $71 \pm 6~M_{\odot}\,{\rm pc^{-2}}$, respectively \citep[KG89a,b,c][hereafter KG91]{kuij91}.
The uncertainties are mainly caused by the measured errors in distance and velocities of the K-dwarfs and the technique of recovering the vertical force $K_z$.

After KG91 there have been various other determinations of the disc surface mass density based on different stellar tracer: \citet[][hereafter FF94]{flyn94} derived a surface mass density of $52 \pm 13~M_{\odot}\,{\rm pc^{-2}}$ for the known disc matter. With Hipparcos data, \citet{korc03} focused $\Sigma_{|z|<350~{\rm pc}} = 42 \pm 6~M_{\odot}\,{\rm pc^{-2}}$, by using red giants; \citet[][hereafter HF04]{holm04} and \citet{bien06} used K giants and red clump stars to estimate the disk surface density again, and the values are
$\Sigma_{|z|<1.1~{\rm kpc}} = 74 \pm 6~M_{\odot}\,{\rm pc^{-2}}$ and $\Sigma_{|z|<1.1~{\rm kpc}} \sim 57 - 79 \pm 6~M_{\odot}\,{\rm pc^{-2}}$, respectively. Recently, \citet[][hereafter G11 and G12]{gar11,garb12}  reviseited the problem, drawing on Jeans equation modelling; they used Markov-Chain Monte Carlo (hereafter MCMC) to marginalize over unknown parameters and data from the literature to obtain an estimate of $\Sigma_{|z|<1.1~{\rm kpc}} = 45.5^{+5.6}_{-5.9}~M{\odot}\,{\rm pc^{-2}}$ for the baryonic disc mass and of $\rho_{\rm DM} = 0.025^{+0.014}_{-0.013}~M_{\odot}\,{\rm pc^{-3}}$.
The rather broad range in published values implies that better data and techniques are needed to improve the measurement of the density distribution of Galactic disk.

In this paper, using the large sample of K-dwarfs observed in SDSS/SEGUE \citep{yann09}, we re-determine the disc surface mass density and mass density of dark matter. In a number of aspects, this work follows KG89b, in particular in constraining the ``vertical force'' $K_z$. However, the paper presents a number of new elements over previous studies in addition to the new large data set. We split the data into abundance selected sub-samples, which provide distinct probes of the same potential; and we simultaneously fit densities and kinematics using MCMC approach. Besides, we explore how the results depend on the functional forms for $K_z$.

This paper is organized as follows. In $\S$~\ref{sec:data}, we describe the SEGUE/SDSS data, in particular we emphasize the spatial selection function, the distance determination, and the sub-samples of similar [Fe/H] and [$\alpha$/Fe]. In $\S$~\ref{sec:determine_lms} we lay out how we determine the tracer number density and vertical velocity dispersion, and practicalities of solving the Jeans equation. The fitting results are shown and discussed in $\S$~\ref{sec:results} and $\S$~\ref{sec:discussion}, with a summary in $\S$~\ref{sec:conclusion}.

\section{DATA}
\label{sec:data}

For our analysis we aim for a sample of kinematics tracers that can be found in disk populations of various ages and metallicities, and which cover distance of 0.2~kpc to 2~kpc from the Sun. In the Sloan Digital Sky Survey \citep[SDSS;][]{york00}, K-dwarfs best satisfy these criteria. The Sloan Extension for Galactic Understanding and Exploration (SEGUE) is a subsurvey of SDSS-II \citep{aba09} which operated from 2005 August to 2008 July, to probe the formation and evolution of our Galaxy. It obtained $ugriz$ imaging of some 3500 deg$^2$ of sky outside of the SDSS-I footprint \citep[][and references therein]{yann09}, with special attention being given to scans of lower Galactic latitudes ($|b| < 35^\circ$) in order to better probe the disk/halo interface of the Milky Way. Overall, SEGUE obtained some 240,000 medium-resolution (R $\sim$ 2000) spectra of stars in the Galaxy, selected to explore the nature of stellar populations from 0.3 kpc to 100 kpc \citep{yann09}. The seventh data release (DR 7) is the final public data release from SDSS-II occurring in October 2008. SDSS-III, which is presently underway, has already completed the sub-survey SEGUE-II, an extension intended to obtain an additional sample of over 120,000 spectra for distant stars that are likely to be members of the outer-halo population of the Galaxy. Data from SEGUE-II have been distributed as part of the eighth public data release (DR 8) \citep{aiha11a}.  The SEGUE Stellar Parameter Pipeline processes the wavelength- and flux-calibrated spectra generated by the standard SDSS  spectroscopic reduction pipeline \citep{stou02}, obtains equivalent widths and/or line indices for more than 80 atomic or molecular absorption lines, and estimates $T_{\rm eff}$, $\log g$, and [Fe/H] through the application of a number of approaches \citep[see][]{lee08a,lee08b,all08,smo11}.

\subsection{K-dwarf selection}
\label{subsec:color_cuts}

KG89b suggested that the tracer star in the present context should have the following properties: 1) they are phase-mixed and in dynamical equilibrium in Galactic potential; 2) they are common in order to satisfy sufficient statistical precision in the result; 3)  they can be found to $|z| > 1.0$~kpc to make sure that the total surface mass density can be measured; 4) the distance of tracer star can be well determined. Therefore, considering present SDSS/SEGUE observations, K-dwarfs are ideal stars which are used as tracer to measure the total disc mass. Compare to the SEGUE G-dwarfs (e.g. \citealt{bovy12a}), they have the advantage that their minimal distance to be included in SDSS/SEGUE is almost two times closer.

\citet{yann09} list the following color and magnitude cuts for identification and targeting of K-dwarf candidates:
\begin{center}
   14.5 $\leq \, r_0 \, \leq$ 19.0, \\
   0.55 $\leq \,(g - r)_0 \, \leq$ 0.75.
\end{center}
In order to get reliable estimates for the `vertical' ($z$) component of each star's velocity, we added two more criteria, the existence of a good proper motion measurement and the existence of spectra in the data base
\begin{center}
   error of proper motion $>$ 0., \\
   $S/N \,>$ 15.
\end{center}
These criteria return 10,925 candidates in DR 8. However, DR 8 does not list [$\alpha$/Fe], which is very important in subsequent analysis for K-dwarf candidates. Besides, proper motions of DR8 above declination $\delta\,>41^\circ$ are worse than those of DR7 because of the astrometric calibration in DR8 \citep{aiha11b}.
Thus, we use the [$\alpha$/Fe] and proper motions from DR 7 for  all of these candidates. To eliminate K-giants, only stars whose $\log g > 4.0$ are taken in the present work, which in the end leaves 9,157 stars. Fig.~\ref{fig:num_den} shows the number density distribution of sample stars in [$\alpha$/Fe] vs. [Fe/H] space.

\subsection{Abundance-selected sub-samples}
\label{subsec:mono}
\begin{figure}
  \centering
 \includegraphics[width=0.85\textwidth]{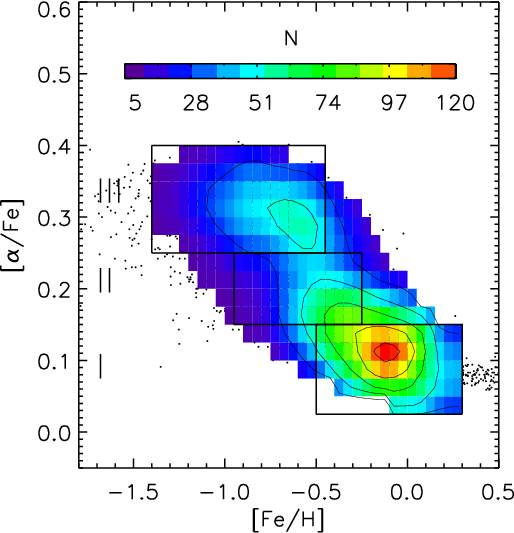}
 \caption{The unweighted number distribution of the 9157 K-dwarfs from SDSS/SEGUE used in our analysis, shown in the [$\alpha$/Fe]-[Fe/H] abundance space, with abundances taken from the SSPP \citep{lee08a,lee08b}. As volume completeness corrections have not yet been taken into account, this distribution overemphasizes the metal-poor, $\alpha$-enhanced stars \citep[see][]{liu12,bovy12a}. Black boxes indicate the three abundance-selected sub-populations. I: metal-rich ([Fe/H] $\in$ [-0.5, 0.3], [$\alpha$/Fe] $\in$ [0., 0.15]); II: intermediate metallicity ([Fe/H] $\in$ [-1.0, -0.3], [$\alpha$/Fe] $\in$ [0.15, 0.25]); III metal-poor ([Fe/H] $\in$ [-1.5, -0.5], [$\alpha$/Fe] $\in$ [0.25, 0.50]).}
 \label{fig:num_den}
 \end{figure}

As shown in Fig.~\ref{fig:num_den}, the distribution of the K-dwarf sample in the [$\alpha$/Fe] $-$ [Fe/H] plane is bi-modal, i.e. a metal-rich, $\alpha$-deficient population and a metal-poor, $\alpha$-enhanced one, as found by \citet{lee11} and \citet{bovy12a} for SDSS G-dwarfs.

\citet{liu12} and \citet{bovy12b} have shown that the kinematics of `mono'-abundance sub-populations are relatively simple. Therefore, we will split our sample of K-dwarfs into subsets that are abundance-selected in the [$\alpha$/Fe] $-$ [Fe/H] plane (see Fig.~\ref{fig:num_den}, black boxes)
 \begin{itemize}
      \item[] $\bullet$ metal-rich: [Fe/H] $\in$ [-0.5, 0.3], [$\alpha$/Fe] $\in$ [0., 0.15]
      \item[] $\bullet$ intermediate metallicity: [Fe/H] $\in$ [-1.0, -0.3], [$\alpha$/Fe] $\in$ [0.15, 0.25]
      \item[] $\bullet$ metal-poor: [Fe/H] $\in$ [-1.5, -0.5], [$\alpha$/Fe] $\in$ [0.25, 0.50]
\end{itemize}
These sub-samples contain 3672, 1416, and 2001 stars, respectively. Of course,
all these sub-population stars move in the same gravitational potential.
We will present the analysis, separating the three sub-populations.
\section{The Galactic potential near the Sun}
\label{sec:determine_lms}
To determine the integral surface mass density and hence measure the Galactic potential near the Sun, with the method described in KG89a, KG89b used K-dwarfs as tracers and assumed that the velocity distribution function of the tracer is only a function of vertical energy: $f_{z}(v_z, z) = f_z(E_z)$.  Adopting the observed vertical tracer density profile $\nu$ and a set of parameterized gravitational potential models $\Phi$, which describe the disc surface density of baryonic matter and the volume density of dark matter halo, KG89b use a Jeans equation which includes $\sigma_{Rz}$-term
\begin{equation}
\label{eq:jeans_kg89}
-\nu\frac{\partial \Phi}{\partial z} = \frac{\partial}{\partial z}(\nu \sigma_{zz}^2) + \frac{1}{R}\frac{\partial}{\partial R}(R \nu \sigma_{Rz}^2),  
\end{equation}
where $\sigma_{ij}^2 = <v_iv_j>$ is the velocity dispersion tensor, and an Abel transform
\begin{equation}
f_z(E_z) = \frac{1}{\pi}\int_{E_z}^{\infty} \frac{-{\rm d}\nu/{\rm d}{\Phi}}{\sqrt{2(\Phi-E_z)}}{\rm d}\Phi
\end{equation}
to predict $f_z(E_z)$ for each $\Phi$. Convolving it with error distribution, they calculated  modeled $f^{\rm mod}_z(v_z,z)$ and compare it with spectroscopic samples to find the best potential parameters. In KG89a's  method, $\sigma_{Rz}$-term was also modeled according to the assumption that the axis ratio of the stellar velocity ellipsoid remains constant in spherical-polar coordinates. 

In our analysis, we will also adopt Jeans equation approach and an assumption from KG89b: the Galactic potential is the sum of contributions by baryonic matter and by dark matter. However, we will predict the vertical tracer density profile and the vertical velocity dispersion simultaneously, and we introduce another assumption: the $\sigma_{Rz}-$term in Jeans equation is neglected (see also G11 and G12). To illustrate the validity of the second assumption, we calculate the velocity dispersion tensor directly to estimate the orders of magnitude of the $\sigma_{zz}-$ and $\sigma_{Rz}-$terms in Eq.~\ref{eq:jeans_kg89} (detail calculations of velocity components are described in \S~3.2). As discussions of the asymmetric drift in \citet[\S~4.8.2 \& \S~4.9.3,][]{bin08} , in a cylindrical coordinate, if assuming that $\sigma_{RR}^2$ and $\sigma_{zz}^2$ both decline with $R$ as $\exp(-R/R_d)$, and the velocity ellipsoid points toward the Galactic center, the $\sigma_{Rz}$-term in Jeans equation can be contained by
\begin{equation}
\left| \frac{1}{R}\frac{\partial}{\partial R}(R\nu\sigma_{Rz}^2) \right| \simeq \frac{2\nu}{R_d}\sigma_{Rz}^2 \lesssim \frac{2\nu z}{R_d} \frac{\sigma_{RR}^2 - \sigma_{zz}^2}{R_{\odot}},
\end{equation}
and the $\sigma_{zz}-$term is on the order of $\nu \sigma_{zz}^2/z_0$, where $z_0$ is the disc scale height. We calculate the average value of $<v_Rv_z>$ to roughly estimate $\sigma_{Rz}^2$, and it is smaller than $\sigma_{zz}^2 = <v_z^2>$  by a factor of $\sim 0.02$ for our present sample stars. $z_0 \ll R_d$, therefore, $\sigma_{Rz}-$term is smaller than $\sigma_{zz}-$term by at least a factor of $2(\sigma_{Rz}^2/\sigma_{zz}^2)\times(z_0/R_d) \lesssim 0.01$. This satisfy the second assumption. It also implies that we can consider the simple problem of solving the Jeans equation for a one-dimensional slab and then to determine gravitational potential and hence density of matter near the Sun ($|R -R_{\odot}| = 1.0~{\rm kpc}$ and $|z| \lesssim 1.5~{\rm kpc}$).  Specifically, the ``vertical'' Jeans equation can be written as \citep{bin08}
\begin{equation}
	\label{eq:vertjeans}
	\frac{\drm}{\drm z} \left[ \nu_{\star}(z) \sigma_z(z)^2 \right]
		= -\nu_{\star}(z) \frac{\drm \Phi(z,R_{\odot})}{\drm z}
\end{equation}
where $\Phi(z,R_{\odot})$ is the vertical gravitational potential in the Sun's vicinity, i.e., we solve this equation for $R = {\rm constant}$. Here, $\nu_{\star}(z)$ is the vertical number density of the tracer stellar population, and $\sigma_z(z)$ the vertical velocity dispersion of exactly those tracers.

With one-dimensional Poisson equation which relates the potential to the disk vertical mass density and the first assumption, in cylindrical coordinates we can get
\begin{equation}
    \label{eq:vertpoisson}
    4 \pi G \rho_{\rm tot} (z,R_{\odot})\,=\, \frac{\drm^2}{\drm z^2 }( \Phi_{\rm disk}(z,R_{\odot})+\Phi_{\rm DM}(z,R_{\odot})),
\end{equation}
therefore
\begin{equation}
\label{eq:rho_tot}
\rho_{\rm tot} (z,R_{\odot})\,=\, \rho_{\rm disk}(z,R_{\odot}) + \rho_{\rm DM}^{\rm eff}(z,R_{\odot})
\end{equation}
where $\rho_{\rm tot}(z,R_\odot)$ is the total mass density of disk components $\rho_{\rm disk}(z,R_{\odot})$ and $\rho_{\rm DM}^{\rm eff}(z,R_{\odot})$, an effective dark matter contribution that includes the circular velocity term. In the Solar neighborhood it can be written as G11
\begin{equation}
\label{eq:rho_rot}
\rho_{\rm DM}^{\rm eff} (z,R_{\odot})= \rho_{\rm DM}(z,R_{\odot}) - \frac{1}{4 \pi G R}\frac{\partial}{\partial R}V_c^2(R_{\odot})
\end{equation}
where $V_c(R_{\odot})$ is the circular velocity at $R_{\odot}$. As discussed in G11, if the disc scale height is much smaller than the dark matter halo scale length, the dark matter density can be assumed to be constant over the range of $|z|$ which is not far from the Galactic mid-plane, that is, $\rho_{\rm DM}(z,R_{\odot}) \simeq \rho_{\rm DM}(R_{\odot})$. Because in the Solar neighborhood the rotation curve is almost flat \citep{bin08,bovy12e}, $\rho_{\rm DM}^{\rm eff} \simeq \rho_{\rm DM}(R_{\odot})$. Therefore, it is possible to introduce the third assumption: the dark matter density $\rho_{\rm DM}$ is taken to be constant within the modeled volume.

Because the first derivative of $\Phi(z,R_{\odot})$ is the gravitational force perpendicular to the galactic plane $K_z(z)$
\begin{equation}
\label{eq:kz}
  K_z (z,R_{\odot}) \equiv -\frac{\drm \Phi(z,R_{\odot})}{\drm z}
\end{equation}
%
Combining Eq.~\ref{eq:vertjeans} and Eq.~\ref{eq:kz}, one obtains 
\begin{equation}
    \label{eq:vertfinal}
    \frac{\drm}{\drm z} \left[ \nu_{\star}(z) \sigma_z(z)^2 \right] = \nu_{\star}(z) K_z(z,R_{\odot})
\end{equation}
To obtain the local mass density distribution from Eq.~\ref{eq:vertfinal}, we adopt a parameterized form of $K_z(z,R_\odot\mid\vec{p})$, where $\vec{p}$ is set from Eq.~\ref{eq:rho_tot} and it contains the integral surface mass density of baryonic matter $\Sigma_{\rm disk} = \Sigma_{\star}+\Sigma_{\rm gas}$ and its scale height $z_h$, and the volume density of dark matter $\rho_{\rm DM}$. Note that we use $\Sigma_{\star}+\Sigma_{\rm gas}$ rather than $\rho_{\rm disk}$, because $\rho_{\rm disk}$ and $z_h$ are degenerate, i.e., one can not constrain $\rho_{\rm disk}$ and $z_h$ separately at the same time.

Then for a given tracer sub-population, we proceed as follows:
\begin{enumerate}
     \item[i)] We determine $\nu_{{\star},{\rm obs}}(z)$ and $\sigma_{z,{\rm obs}}(z)$ in bins of $|z|$ for each of the three tracer populations directly from the observations.
     \item[ii)] We model $\nu_{\star}(z)$ as a simple exponential \citep{bovy12a} of unknown scale height $\vec{h}_z$. Here $\vec{h}_z = [h_{z,1}, h_{z,2}, h_{z,3}]$ correspond to the metal-rich, inter-intermediate metallicity, and metal-poor sub-samples, respectively. 
     \item[iii)] We put $\nu_{\star}(z\mid\vec{h}_z)$ and $K_z(z,R_\odot\mid\vec{p})$ into Eq.~\ref{eq:vertfinal} and solve for $\sigma_z(z\mid\vec{p},\vec{h})$.
     \item[iv)] For each [$\vec{p},\vec{h}_z$], we compare $\nu_{\star}(z\mid\vec{h}_z)$ and $\sigma_z(z\mid\vec{p},\vec{h}_z)$ with $\nu_{{\star},{\rm obs}}(z)$ and $\sigma_{z,{\rm obs}}(z)$ by calculating the likelihood function $\mathcal{L}(\sigma_z,\nu_\star\mid\vec{p},\vec{h}_z)$.
     \item[v)] We use a MCMC technique to sample the likelihood of the data, given [$\vec{p},\vec{h}_z$]  in order to find the best parameters.
\end{enumerate}

In this modeling, we parameterize $K_z$ instead of the gravitational potential. For any $K_z$ function form, the gravitational potential can then be derived by integration of $K_z$. 
Here we will describe the details step by step.

\subsection{Tracer number density}
\label{sec:density_law}
\subsubsection{Distance estimates and coordinate system}
\label{subsubsec:dis}
To obtain $z$ and $v_z$ of our sample stars in the Solar vicinity, we need to estimate their distance first. Given the de-reddened color indices ($(g-r)_0$ and $(g-i)_0$), the de-reddened $r$-band magnitude (in SDSS/DR8, the extinction corrections in magnitudes are computed following \citet{schl98}), and metallicity, the absolute magnitude is estimated by fitting to fiducial color-magnitude relations, calibrated through star cluster spectroscopy. The fiducial sequences for $(g-r,\,r)_0$ and $(g-i,\,r)_0$, based on YREC+MARCS isochrones, as described in \citet{an09}, are adopted. Then it is straightforward to determine the distance $D=10^{\frac{r_0-M_r}{5}+1}\,{\rm pc}$.

Such distances were estimated by fitting the fiducials for the two colors above separately, yielding an average difference and standard deviation among these two distance estimates of $0.006\pm0.077$~kpc.
We use the average distance and the standard deviation from color-magnitude (hereafter CM) diagrams of  $(g-r,\,r)_0$ and $(g-i,\,r)_0$ as the mean distance and its error, respectively.

With $D$ estimated as above and observed Galactic
longitude and latitude($l,\,b$), we get the stars' position in Galactic cylindrical coordinates. We adopt a Galactocentric cylindrical coordinate system in which $\hat{\textbf{z}}$-axis towards the North Galactic Pole ($b\,\backsimeq\,90 ^\circ$), that is,
\begin{equation}
\begin{aligned}
       & R=\sqrt{D^2\cos^2(b)-2DR_{\sun}\cos(b)\cos(l)+R^2_{\sun}} \\
       & \theta = \arctan \left[\frac{D\cos(b)\cos(l)-R_{\sun}}{D\cos(b)\sin(l)} \right]\\
       & Z=D\sin(b) \\
\end{aligned}
\end{equation}
where $R_{\odot}\,=\,8.0$ kpc is the adopted distance to the Galactic center \citep{reid93}. We aim to obtain the total surface mass density distribution enclose within a slab of height $|z|$ in the Solar neighborhood, as in G12 and \citet[][hereafter BT12]{bovy12d}. 

We will treat the dynamics as 1-D problem ($\hat{\textbf{z}}$-direction) in the following. This approximation in our analysis can be justified by the limited radial extent of our sample, with a median $|R-R_{\odot}|=0.4$~kpc, and by its limited vertical extent, with a median $|z|=0.8$~kpc. Therefore, $|R-R_{\odot}|$ and $|z|$ are $\ll R_{\odot}$ and the neglect of any radial gradients and the velocity ellipsoid tilt should be good approximations.

\subsubsection{Selection function}
\label{subsubsec:sel_fun}

Any dynamical analysis need accurate knowledge of the spatial distribution of the kinematic tracer. Therefore, we need not only a well-defined spectroscopic sample of K-dwarfs with known fluxes and distances (or heights above the mid-plane) estimates, we also need to understand their selection function \citep{bovy12c}, which is what we lay out there.

In this coordinate system, each sub-sample can be divided in eight bins according to $R$ and $\Delta R = R_{j+1}-R_{j}=500$ pc. In each $\Delta R$ range, $R$ is considered as a constant, and each sub-sample is also divided in nine bins in $\hat{\textbf{z}}$ direction. The width of the $z$ bin, $\Delta z \equiv z_{i+1} - z_i$, is $\Delta z = 100\,{\rm pc}$. Here $j$ and $i$ are indices of $R$ and $z$ bins, respectively. In each $z$-$R$ box
\begin{equation}
       \nu_{\star,s} (R_j,z_i)\,=\,\frac{N_s(R_j,z_i)}{V_{{\rm eff},s}(R_j,z_i)}
\end{equation}
where $s$ denotes the different abundance-dependent sub-populations, $N_s(R_j,z_i)$ and $V_{{\rm eff},s}(R_j,z_i)$ are respectively the star number and the effective volume which is corrected by selection function in a ($R_j$,$z_i$) box. Because every targeted star of a given $(g-r,\,r)_0$ has a possibility to be in one particular sub-sample, each line of sight is a part of the \emph{search} volume for every abundance sub-sample. In our effective volume calculation, all lines-of-sights of all K-dwarfs are included, i.e.
\begin{equation}
       V_{{\rm eff},s}(R_j,z_i)\,=\,\sum_{q=1}^{n_s}V_{{\rm eff},q}
\end{equation}
where $n_s$ is the total number of lines-of-sights of a given sub-sample and $V_{{\rm eff},q}$ is the \emph{search} volume of each line-of-sight.

It is in the calculation of the effective volume that the selection function enters explicitly. Some aspects of the SEGUE selection function are obvious. The apparent magnitude range brackets the possible distances of K-dwarfs, and the most nearby stars will be preferentially the brightest and coldest stars. At a given $r_0$ and $(g-r)_0$, the effective survey volume is smaller for lower metallicity (hence less luminous) stars. Moreover, the SDSS targeting strategy implies that stars at lower latitude have a smaller probability of ending up in the spectroscopic sample, since a fixed number of targets is observed along each line of sight. In general, we need to derive the selection function, which gives the probability that a star of given $M_r$, $(g-r)_0$, [Fe/H] ends up in the samples, as a function of $D$ and $(l,b)$

\begin{figure}
  \centering
 \includegraphics[width=\textwidth]{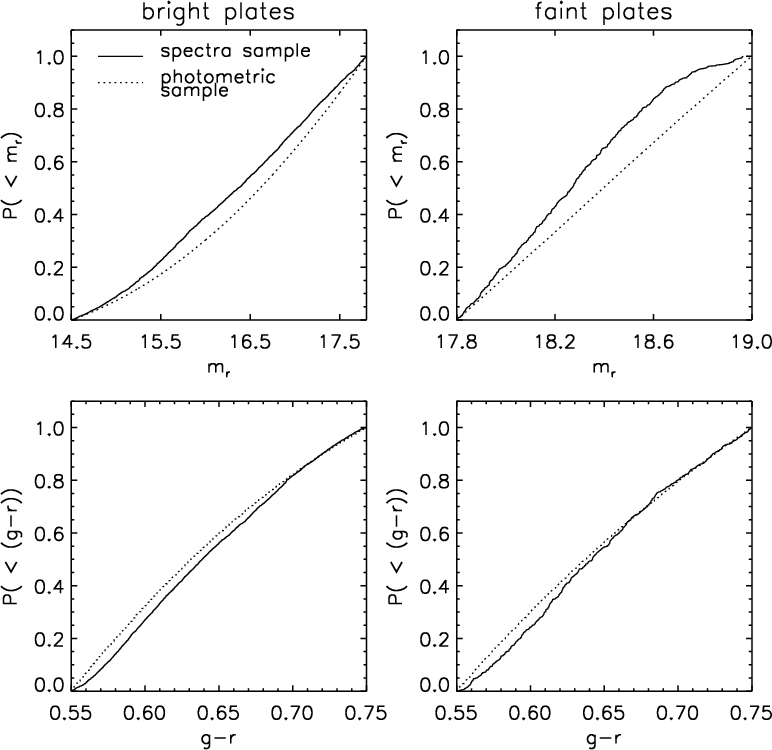}
 \caption{Spectroscopic success rate for K-dwarfs in SDSS/SEGUE, used to estimate the selection function (\S~\ref{subsubsec:sel_fun}). The panels show the comparison of the distribution of the K-dwarfs that have good spectra from SEGUE (solid line) with the distribution of the presumably complete distribution of all photometrically detected point-source in the plate that satisfy the color-magnitude selection criteria.}
 \label{fig:sel_fun}
 \end{figure}

Each SEGUE pointing which corresponds to an angular on-the-sky radius of 1.49\dgr is observed with two plates, a SEGUE bright plate that targets stars with $14.0 < m_r < 17.8$ and faint plate that targets stars with $17.8 < m_r < 20.1$ \citep{yann09}.
In each plate, the distribution of spectroscopic sample in a CM box that satisfies the selection criteria is $n_{\rm spec}(g-r,\,r)$, while the distribution of all photometric stars within the same plate that satisfy the same CM cuts is $n_{\rm photo}(g-r,\,r)$. Integration of these distribution over all CM cuts results in $N_{\rm spec}$ and $N_{\rm photo}$, leading to a plate weight
\begin{equation}
\label{eq:p_w}
W_{{\rm plate},k} \equiv \frac{N_{{\rm spec},k}}{N_{{\rm photo},k}}
\end{equation}
where $k$ is the index of plate. This plate weight should be a part of selection function. Fig.~\ref{fig:sel_fun} presents the cumulative distributions of spectroscopic and photometric samples in the same selection criteria. It is clear that the selection weight for faint plates has an apparent magnitude dependence because of the dependence on signal-to-noise ratio of the probability of obtaining a good spectrum but no strong color dependence.
Therefore, we make an approximation for this selection function, that is, it is a function of $r_0$.

This function is defined for all distance moduli, which is one for distance moduli that correspond to apparent magnitude insides of plates' magnitude range and zero for other situation. We combine the selection function in to the calculation of $V_{{\rm eff},q}$, that is, the average metallicity of one particular sub-sample $<$[Fe/H]$>$ is adopted to estimate the possible absolute magnitude, $M_q$, distance, $D_q$, and height above mid-plane, $z_q$, of a given ($r,\,g-r$). Then
\begin{equation}
      V_{{\rm eff},q}\,=\,\frac{\pi}{3}\cdot W_{{\rm plate},q}\cdot W_q\cdot\frac{\sin(\theta)}{\sin(b_{q})}\cdot \frac{1}{2}[\cot(b_{q} - \theta) - \cot(b_{q} + \theta)]\cdot(z_{\rm upper}^3-z_{\rm lower}^3)
\end{equation}
where $\theta\,=\,1.49$\dgr is the diameter of SEGUE (and SDSS) spectroscopic plate \citep{yann09}, and $W_{{\rm plate},q} = W_{{\rm plate},k}$ in a given plate. Besides,
\begin{equation}
      W_q\,=\,\left\{
      \begin{array}{ll}
      1,& {\rm for}\,\,z_{i} \leqslant z_q \leqslant z_{i+1} \\
      0,& {\rm otherwise}
      \end{array}
      \right.
\end{equation}
\begin{eqnarray}
      z_{\rm upper}&=&{\rm Min}(z_{i+1}, D_{r,{\rm max}}*\sin(b_q)) \\
      z_{\rm lower}&=&{\rm Max}(z_{i}, D_{r,{\rm min}}*\sin(b_q))
\end{eqnarray}
\begin{eqnarray}
      D_{r,{\rm min}}&=&10^{\frac{r_{\rm min}-M_q}{5}+1} \,{\rm pc}\\
      D_{r,{\rm max}}&=&10^{\frac{r_{\rm max}-M_q}{5}+1} \,{\rm pc}
\end{eqnarray}
where $r_{\rm min}$ and $r_{\rm max}$ are the magnitude limits of each plate. For SEGUE bright plate, $r_{\rm min}=14.0$ and $r_{\rm max}=17.8$, and for SEGUE faint plate, $r_{\rm min}=17.8$ and $r_{\rm max}=20.1$ \citep{yann09}. In each $z_i$ bin, the error bar of the tracer number density arises from the star count Poisson variance and it is estimated by means of Monte Carlo bootstrap ping \citep[$\S$~15.6 of ][]{pre07}.

Fig.~\ref{fig:tracer_den} presents the resulting ($R$,$z$) map of the tracer number density for each sub-population. From top to bottom, those plots correspond metal-rich, intermediate metallicity, and metal-poor sub-sample; from left to right, plots represent effective star number, effective volume, and natural logarithm of number density in each $z$-$R$ box. In this figure, one can see that the scale-height of metal-rich sub-sample is shorter, but the scale-length of the same sub-population is longer. This result is similar as the studies of Galactic structure of \citet{bovy12c}. In each $z_i$ bin, we calculate an average number density over all $R$, and then to get the vertical profile of the tracer number density.

\subsection{Vertical velocity dispersions}
\label{subsec:vvd}

\begin{figure}
 \centering
 \includegraphics[scale=0.75]{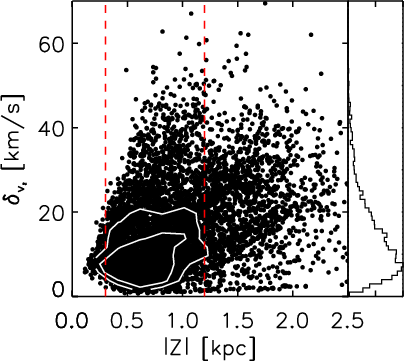}
 \caption{\textbf{The distribution of the error of vertical velocity dispersion. Stars between two red dashed lines are samples we used in the present work. The histogram represents the distribution of the used stars and contours mean 68\% and 95\%  confidence intervals of this distribution. }}
 \label{fig:v_z_error}
 \end{figure}

\begin{figure}
  \centering
 \includegraphics[width=\textwidth]{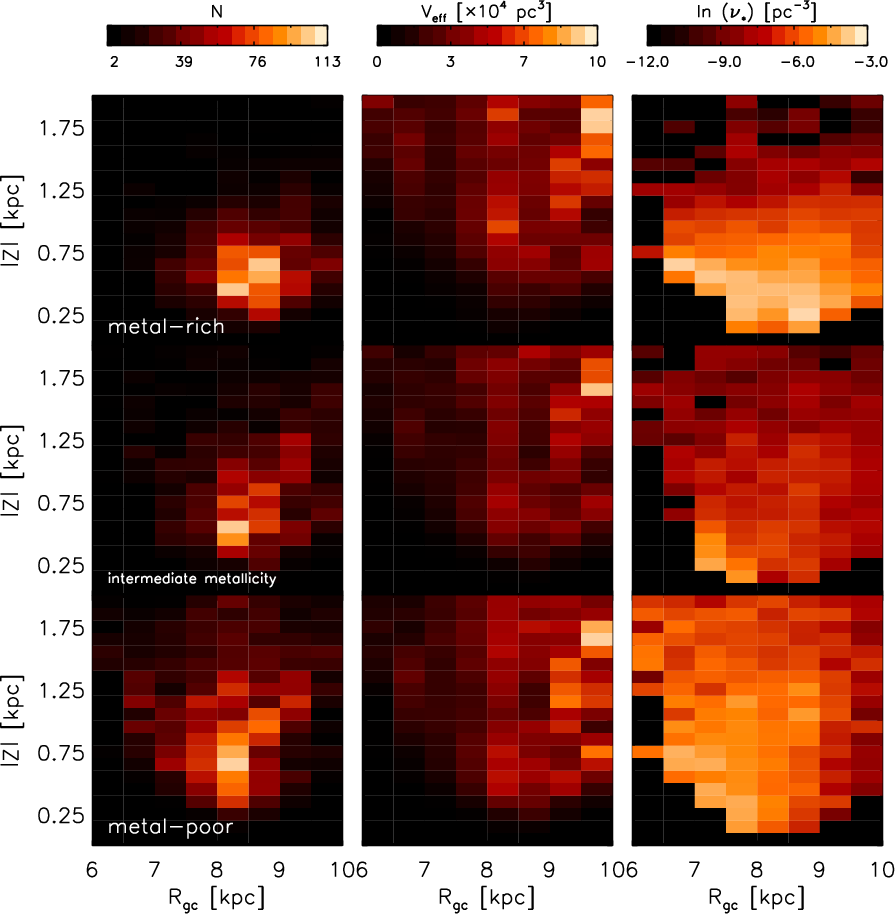}
 \caption{Derivation of the spatial tracer number density for each abundance-selected sub-sample. From top to bottom, plots show the metal-rich, intermediate metallicity, and metal-poor sub-populations, respectively. From left to right, the panels show the  actually detected number of stars as a function of galactocentric distance and height above the plane, the effective survey volume, and the implied number density (see \S~\ref{sec:density_law}).}
 \label{fig:tracer_den}
 \end{figure}

The basic kinematic measurements for each stars are its proper motions measured along Galactic longitude, and latitude $\vec{\mu}_{l,b}$, and the radial velocity $v_{rad}$. From this we calculate the spatial velocity of stars and convert it to the  Galactic center (GC) by adopting a value
of 220~km/s for the local standard of rest ($V_{LSR}$) and a solar motion of (+10.0, +5.2, +7.2)~km/s in (\textbf{$U$, $V$, $W$}), which are defined in a right-handed Galactic system with \textbf {$U$} pointing toward the Galactic center, \textbf{$V$} in the direction of rotation, and \textbf{$W$} toward the north Galactic pole \citep{dehn98}.  In the Galactocentric cylindrical coordinate system adopted in the present study, $W$ is the vertical velocity $v_z$. The error of $v_z$, $\delta_{v_z}$, is mainly from propagation of observed errors of $\vec{\mu}_{l,b}$ and $v_{rad}$ and the error of distance. Fig.~\ref{fig:v_z_error} shows the distribution of $\delta_{v_z}$, which shows that the median error is $\sim$ 13 km/s, and the errors is smaller than 25 km/s for 95\% of the sample stars. Therefore, it is important that the velocity error of the individual stars are properly folded into the estimate of the velocity dispersion. To estimate $\sigma_z(z,{\rm [Fe/H],[\alpha/H]})$, one can therefore not simply calculate the standard deviation of the observed $v_{z,i}$ values for each $z$ bin of each sub-population. Instead, we use maximum likelihood technique described in \citet[][Appendix A]{van06} to estimate the intrinsic velocity dispersion, corrected for all individual velocity errors.


In each $z$ bin, the intrinsic velocity distribution of the stars is assumed as
$\mathcal{L}(v_z)$. Each stellar velocity $v_{z,i}$ in this bin is the product of $\mathcal{L}(v_z)$ convolved with a delta function which is broadened by the observed uncertainties $\delta_{v_{z,i}}$, and integrated over all velocities. For all $N$ stars in the bin, the likelihood can be defined by
\begin{equation}
\label{eq:v_like}
\mathscr{L}(\bar{v}_z,\sigma_z,\ldots) = \prod_{i}^{N}\int_{-\infty}^{+\infty}\,\mathcal{L}(v_z)\frac{{\rm e}^{-\frac{1}{2}\left( \frac{v_{z,i}-v_z}{\delta_{v_{z,i}}}\right)^2}}{\sqrt{2\pi} \delta_{v_{z,i}}}{\rm d}v_{z}
\end{equation}
where $\mathscr{L}$ is a function of mean velocity $\bar{v}_z$, velocity dispersion
$\sigma_z$ and possible higher-order velocity moments. The velocity distribution $\mathcal{L}(v_z)$ can be recovered by maximizing the likelihood $\mathscr{L}$.
To describe the velocity distribution, we use a parameterized Gauss-Hermite (GH) series \citep{van93,ger93} because it has the advantage that it only requires the storage of the velocity moments ($\bar{v}_z,\,\sigma_z,\,H_3,\,H_4,\ldots$) instead of the full velocity distribution and the convolution of Eq.~\ref{eq:v_like} can be carried out analytically. For more details on the forms of GH series, we refer the reader to \citet[Sec.~2.2 and Appendix A,][]{van93}.
This makes it feasible to apply the method to a large number
of discrete measurements and to estimate the uncertainties on the extracted velocity moments by means of the Monte Carlo bootstrap method \citep[$\S$~15.6 of ][]{pre07}.

\begin{figure}
 \centering
 \includegraphics[width=\textwidth]{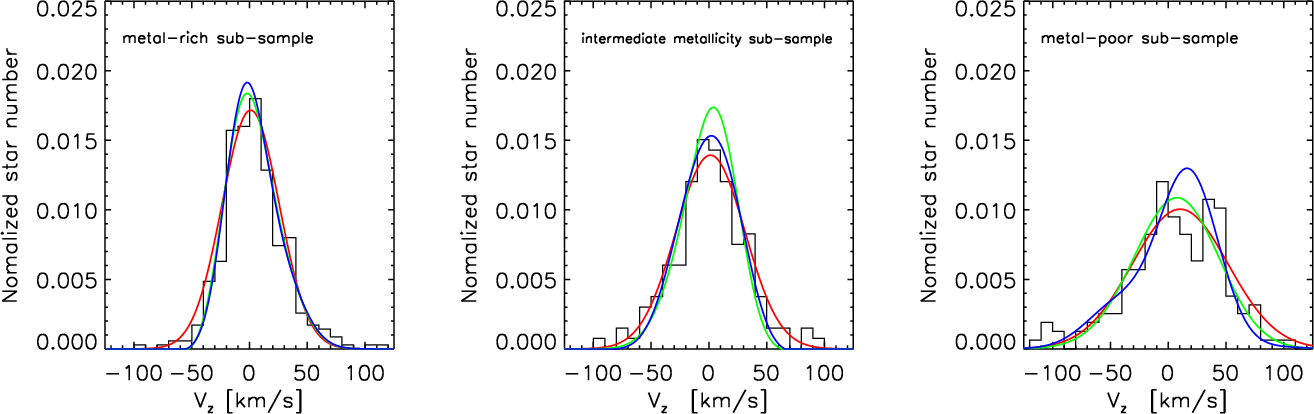}
 \caption{Example fits of the vertical velocity dispersion, shown for three abundance selected samples in the bin 600~pc$ < |z| <$ 700~pc. For all panels, the histograms show the observed vertical velocity distributions, the red solid curves are Gaussian fits without considering observed errors, and the green and blue solid curves are three and four moments Gauss-Hermite accounting for the observational errors. Without taking the individual errors into account overestimates the vertical velocity dispersion by about 2 - 4~km/s.}
 \label{fig:vvd_sample}
 \end{figure}

Fig.~\ref{fig:vvd_sample} shows an example of a vertical velocity dispersion fit in $600\,{\rm pc} < |z| < 700\,{\rm pc}$ for the three sub-populations. This figure shows comparisons of different fitting methods. Red solid curves are the results of Gaussian fits without considering observed errors. Green and blue curves represent three and four moments GH fits using the method described above. For all $|z|$ bins of all sub-populations, the vertical velocity dispersion is overestimated by 2 - 4~km/s when observed errors are not taken into account.

For the metal-rich, intermediate metallicity, and metal-poor sub-samples, the measured vertical velocity dispersion $\sigma_{z,{\rm obs}}(z)$ are in the ranges 17 $-$ 26~km/s, 25 $-$ 37~km/s, and 35 $-$ 42~km/s, showing a slightly increase with $|z|$. Moreover, $\sigma_{z,{\rm obs}}(z)$ also increases with the decline of metallicity and the increasing of [$\alpha$/Fe] (see Fig.~\ref{fig:vertical_disp}, filled circles). The most immediately comparable analysis of the vertical kinematics in abundance-selected sub-samples is that of the SEGUE G-dwarfs \citep{liu12,bovy12c}. The nearly isothermal dispersions in the mono-abundance bins in \citet{bovy12c} that correspond to our metal rich sample range from 15 $-$ 25~km/s; those that correspond to our intermediate bin range from 25 $-$ 35~km/s, and those in the metal-poor bin from 40 $-$ 50~km/s. This is in good agreement with our K-dwarf kinematics, except perhaps for the metal-poor bin. However, a quantitative comparison is complicated by the fact that the velocity  dispersions expected for  populations in broader abundance bins depend on the abundance distribution within that metallicity range \citep[see][]{bovy12c}


\subsection{Functional forms for $K_z$}
\label{subsec:kz_assum}

Given the observed tracer number density and vertical velocity dispersion, we investigate different parameterized $K_z(z)$ forms to solve Eq.~\ref{eq:vertjeans} and derive the corresponding expression for $\sigma_z(z)$.
$K_z(z)$ has contributions from both the baryonic components and the dark matter.
According to the discussion of KG89b, all $K_z(z)$ forms should satisfy the following properties: (1) the halo and disk contributions are degenerate when $|z|\,\ll$ scale height; (2) the disk contribution can be approximated by a thin mass sheet when $|z|\,\gg$ scale height. We considered five different parameterized forms of $K_z(z)$, two of which we discuss below, while the other three are given in the Appendix,
\begin{enumerate}
  \item[i)] {\bf KG89 model}
       \begin{equation}
       \label{eq:kz_kg89}
             K_z = 2 \pi G \left[\Sigma_{\star} \frac{z}{\sqrt{z^2+z_h^2}}+\Sigma_{\rm gas} \right]+
             4 \pi G \rho_{\rm DM} z
       \end{equation}
  \item[ii)] {\bf Exponential model}
       \begin{equation}
       \label{eq:kz_exp}
             K_z = 2 \pi G \left\{ \Sigma_{\star} \left[1-\exp(-\frac{z}{z_h})\right]+\Sigma_{\rm gas}\right\}+
             4 \pi G \rho_{\rm DM} z
       \end{equation}
\end{enumerate}
Heres $\Sigma_{\rm gas} = 13.2$~$M_{\odot}{\rm pc}^{-2}$ is the surface mass density of gaseous components which is taken from the disc mass model of \citet{flyn06}. 


As we stated before, the trace number density profile of each sub-population is modeled by single exponential,
\begin{equation}
\label{eq:num_den}
  \nu_{\star,s} (z) = \nu_{0,s}\times\exp\left(-\frac{z}{h_{z,s}}\right)
\end{equation}
where $s$ is the index of sub-population. For the exponential model, the corresponding vertical dispersion profile of each sub-population can be derived analytically:
\begin{equation}
 \label{eq:sigma_exp}
    \sigma_{z,s}^2 (z)= 2 \pi G h_{z,s}\left\{\Sigma_{\star}\left[1-\frac{z_h}{h_{z,s}+z_h}{\rm exp}(-\frac{z}{z_h})\right]+ \Sigma_{\rm gas}\right\}+
     4 \pi G \rho_{\rm DM} h_{z,s}(z+h_{z,s})
\end{equation}
For the other $K_z$ forms we infer $\sigma_{z,s}$ through numerical integration.

Then it is possible to fit $\sigma_{z,s}(z)$ and $\nu_{\star,s}(z)$ simultaneously and obtain parameters described below.

\subsection{Obtaining the PDFs for the model parameters}
\label{subsec:para_fit}
The predictions for $\sigma_{z,s}(z\mid\vec{p},\vec{h}_z)$ are based on six parameters (i.e. $\vec{p}$=[$\Sigma_{\star}$, $z_h$, $\rho_{\rm DM}$] and $\vec{h}_z$ = $[h_{z,1}, h_{z,2}, h_{z,3}]$) for each $K_z(z, R_\odot)$ model and the vertical tracer populations, which may have considerable covariance. At the same time, we have a parameterized model for the vertical tracer density profile $\nu_{\star}(z\mid\vec{h}_z)$, and we want to derive constraints on the gravitational force, marginalized over $\vec{h}_z$. To explore the parameter space efficiently and to account for parameter degeneracies, we use a MCMC approach to sample the likelihood function, which provides a straightforward way of estimating the probability distributions (PDF) for all the model
parameters and their degree of uncertainty \citep[$\S$~15.8 of ][]{pre07}.

The estimates $\sigma_z(z)$ and $\nu_\star(z)$ in each $|z|$ bin are independent, therefore, the logarithm of the likelihood of the data given the parameters $[\vec{p},\,\vec{h}_z]$ can be written as
\begin{equation}
\begin{aligned}
      \label{eq:like_chi2}
      \ln\mathcal{L}= & -\sum_{m=1}^{M}\ln(2\pi \epsilon_{\sigma_{z,{\rm obs}}}\, \epsilon_{\nu_{\star,\rm obs}})_m\,- \\
       & \sum_{m=1}^{M}\frac{1}{2}\left[\frac{\sigma_z(\vec{p},\vec{h}_z)-\sigma_{z,{\rm obs}}}{\epsilon_{\sigma_{z,{\rm obs}}}}\right]^2_m - \sum_{m=1}^{M}\frac{1}{2}\left[\frac{\nu_{\star}(\vec{h}_z)-\nu_{\star,{\rm obs}}}{\epsilon_{\nu_{\star,\rm obs}}}\right]^2_m
\end{aligned}
\end{equation}
where $M$ is the total number of $|z|$ bin, and $\epsilon_{\sigma_{z,{\rm obs}}}$ and $\epsilon_{\nu_{\star,\rm obs}}$  are the errors of observed $\sigma_{z,{\rm obs}} $ and $\nu_{\star,{\rm obs}}$, respectively. 

In order to reduce the number of parameters by eliminating $\nu_{0,s}$ from Eq.~\ref{eq:num_den}, we calculate $\nu_{0,s}$ for each sub-population by
\begin{equation}
 \label{eq:nu0_cal}
 \frac{\partial \chi^2(\nu)}{\partial \nu_{0,s}} = 0,
\end{equation}
yielding
\begin{equation}
\label{eq:nu0}
  \nu_{0,s} = \left. \sum_{m=1}^M \frac{\exp(-z_m)}{\epsilon_{\nu_{\rm obs},s,m}}\nu_{{\rm obs},s,m} \middle/ \sum_{m=1}^{M}\frac{\exp(-z_m)}{\epsilon_{\nu_{\rm obs},s,m}}\exp\left(-\frac{z_m}{h_{z,s}}\right) \right. .
\end{equation}

All other parameters are estimated by running a MCMC chain of typically 50,000 steps, after a burn-in period. In the Markov chain, the parameters of the gravitational model and scale heights of tracer densities [$\vec{p}_n$, $\vec{h}_{z,n}$] are chosen from a multivariate normal distribution of [$\vec{p}_{n-1}$, $\vec{h}_{z,n-1}$], with the covariance of [$\vec{p}$, $\vec{h}_z$] chosen to yield a total acceptance rate in the MCMC Chain of about $20\%\sim30\%$. We use the results of all steps to represent [$\vec{p}$, $\vec{h}_z$] and their distributions, by counting the rejected steps as repeat ones. After sampling the PDF for parameters for each $K_z$ model, $\chi^2_{\rm tot}$ between the observed values and model predicted ones ($\sigma_{z,{\rm mod}}$ and $\nu_{\star,{\rm mod}}$) is calculated to explore how the models form the different $K_z$ families compare:
\begin{equation}
\begin{aligned}
\label{chi2_compare}
\chi^2_{\rm tot} & = \chi^2_{\sigma_z} + \chi^2_{\nu_{\star}} \\
                 & = \sum_{s=1}^3\left[\sum_{m=1}^M\left(\frac{\sigma_{z,s,{\rm mod}} - \sigma_{z,s,{\rm obs}}}{\epsilon_{\sigma_{z,s,{\rm obs}}}}\right)^2_m+\sum_{m=1}^M\left(\frac{\nu_{s,{\rm mod}}-\nu_{s,{\rm obs}}}{\epsilon_{\nu_{s,{\rm obs}}}}\right)^2_m\right],
\end{aligned}
\end{equation}

\section{Results}
\label{sec:results}

\begin{figure}
 \centering
 \includegraphics[width=\textwidth]{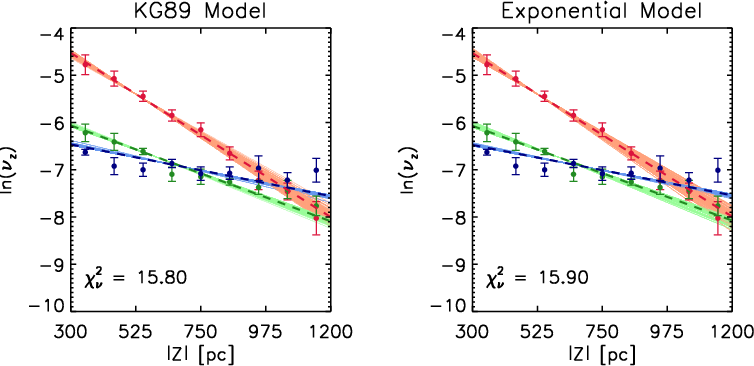}
 \caption{The derived vertical density profiles for the three abundance-selected sub-samples, with the model predictions from two of our parameterized models for the of $K_z$ force profiles (the KG89 and exponential model, respectively; see Eq.~\ref{eq:kz_kg89} \& ~\ref{eq:kz_exp}). Filled circles are values estimated directly from the observations, dashed lines are model predicted values and shadows are the 68.3\% errors in the recovered value of parameters. Red, green, blue symbols correspond metal-rich, intermediate metallicity, and metal-poor sub-population, respectively. $\chi^2_{\nu}$ is calculated from the observed $\nu_{\star,{\rm obs}}$ and modeled $\nu_{\star,{\rm mod}}$ (Eq.~\ref{chi2_compare}).}
 \label{fig:vertical_density}
 \end{figure}

\begin{figure}
 \centering
 \includegraphics[width=0.9\textwidth]{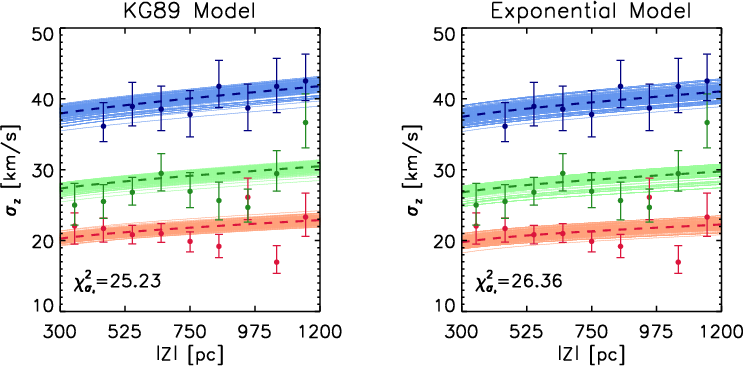}
 \caption{Analogous Figure to Fig.~\ref{fig:vertical_density}, but for vertical velocity dispersions.}
 \label{fig:vertical_disp}
 \end{figure}

The above procedure results in PDFs for the $K_z$ and $\nu_{\star}$ parameters, which are presented in Figs.~\ref{fig:vertical_density} $-$ \ref{fig:dm_pdf}. The best fitting results and recovered parameters for all $K_z$ gravitational force law models are listed in Table ~\ref{tab:result}.

\subsection{Vertical tracer density profiles}

Fig.~\ref{fig:vertical_density} shows the fits to the vertical tracer density profiles. Filled circles are values derived directly from observations, which
are the average value of $R$ columns ($7~{\rm kpc} \leqslant R \leqslant 9~{\rm kpc}$) of the right panel of Fig.~\ref{fig:tracer_den}.
Dashed lines are the predictions for $\nu_{\star,s}(z)$ (Eq.~\ref{eq:num_den}) for two different $K_z$ models, and the shaded regions represent the samplings of the last 200 steps in the MCMC chain.
Red, green, and blue symbols represent metal-rich, intermediate metallicity, and metal-poor sub-population, respectively. Each sub-population has a simple and single exponential structure, and the scale height is increasing with the decline of metallicity \citep[as seen in][]{liu12,bovy12c}.

The joint parameter PDFs show that the tracer scale height, $\vec{h}_z$ is independent of the $K_z$ model parameters (see Fig.~\ref{fig:sh_pdf}). The scale-heights we find for different abundance-select tracer sub-populations are 230 - 260 pc, 450 - 510 pc, and 800 - 1000 pc (see Table~\ref{tab:result}), for all $K_z$ families. The minor differences caused by different models are within the 68\% central region of the $h_z$ PDF.

These findings are in accord with recent results by \citet{bovy12c} and \citet{liu12}. In their work, they adopted G-dwarfs from SDSS/DR7 to analysis the relationship between the distribution of Galactic disc stars and their metallicity in each narrow [$\alpha$/Fe]-[Fe/H] box. They concluded that the scale-height increases continually from 200 pc to 1200 pc with the decline of metallicity.

\subsection{Vertical dispersion profiles}
As discussed in \S~\ref{subsec:vvd}, the $\sigma_{z,{\rm obs}}(z)$ shown in Figure 6, have already been corrected for the velocity errors of the individual stars within each $|z|$ bin.
In these plots, we also summarize $\sigma_{z,{\rm mod}}(z)$ predicted by different $K_z$ models. The results reflect simultaneous fit to all three sub-populations. If the $\sigma_{z,{\rm obs}}(z)$ is fitted for each sub-population separately, there is some slight tension between the fits to the individual sub-populations and the simultaneous fit. The difference of $\sigma_{z,{\rm mod}}(z)$ given by different models is smaller than the error of observed $\sigma_{z,{\rm obs}}(z)$, as in the case of the scale-height predictions of vertical number density. That is, for the three sub-population, $\sigma_{z,{\rm mod}}(z)$ is in the range of 16 - 20~km/s, 25 - 29~km/s, and 35 - 43 km/s (see Table ~\ref{tab:result}), respectively.

We note again, that previous studies of $K_z(z)$ by KG89a,b,c, HF04, \citet{sieb03}, G12 and BT12 had not, or could not, split their tracer samples a priori into abundance sub-bins with nearly isothermal $\sigma_z(z)$ profiles.
In samples with a wide metallicity range
$\sigma_z(z)$ will rise with $|z|$  simply because the mix of mono-abundance populations will change. To disentangle a rising $\sigma_z(|z|)$ due to a dark matter halo from the effect of population mixing, these studies had to model the metallicity distribution, which turned out to be an important source of systematic errors.
Fitting simultaneously to populations that have been split {\it a priori} by their abundances, as done here, reduces this source of systematics.

\subsection{The vertical gravitational force $K_z$}

\begin{figure}
 \centering
 \includegraphics[width=0.9\textwidth]{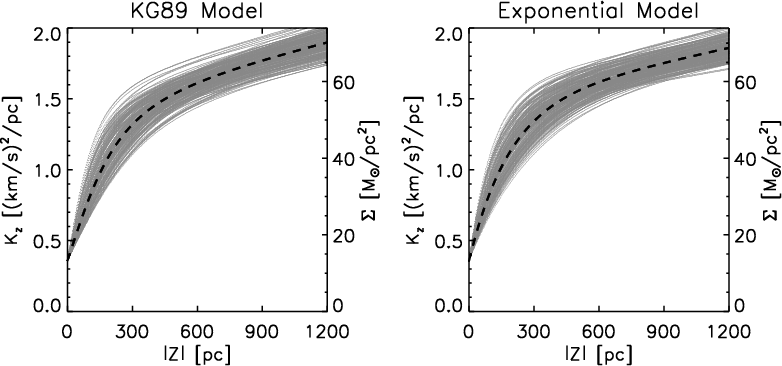}
 \caption{The vertical dependence of the $K_z$ force of our two fiducial models, which include the stellar disk, a (thin) gaseous disk and a dark matter halo term. Left is the KG89 model (Eq.~\ref{eq:kz_kg89}) and right is the exponential model (Eq.~\ref{eq:kz_exp}). In both panels, the dashed fat line shows $K_z(z)$ for the most likely parameters, and the band of grey points show a $1\sigma$ sampling of the PDF for $K_z$.}
 \label{fig:kz_force}
 \end{figure}

\begin{figure}
 \centering
 \includegraphics[width=0.9\textwidth]{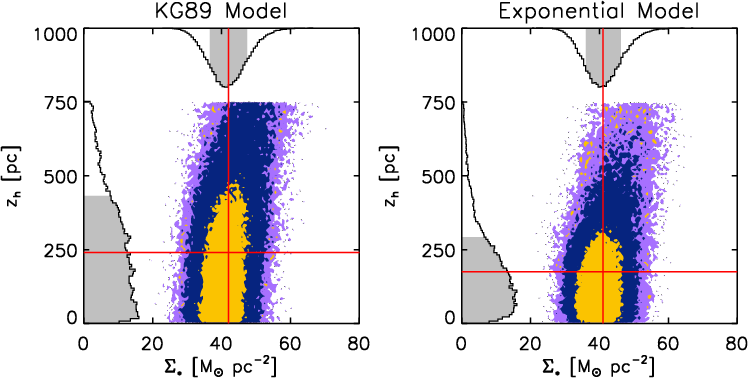}
 \caption{The PDF of model parameters $\Sigma_{\star}$ and $z_h$ for the models in Eq.~\ref{eq:kz_kg89} \& ~\ref{eq:kz_exp}. On the left, the parameters for the KG89 model, on the right for the exponential model. Yellow, blue, and purple shades are 68\%, 95\%, and 99\% confidence region, gray histograms are the marginalized PDFs for the individual parameters, and red lines represent the most likely value of each parameters.}
 \label{fig:para_pdf}
 \end{figure}

\begin{figure}
 \centering
 \includegraphics[width=0.9\textwidth]{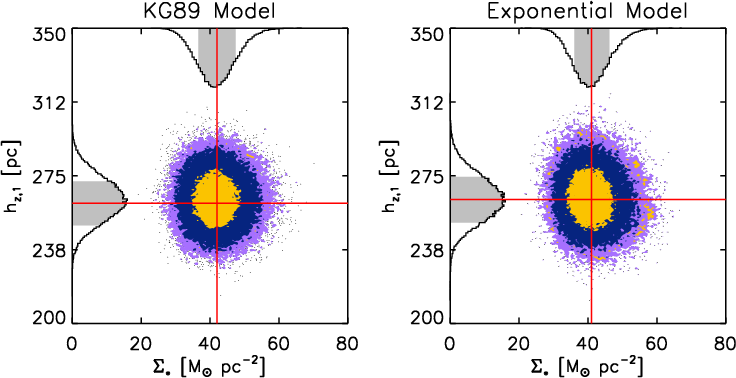}
 \caption{Analogous Figure to Fig.~\ref{fig:para_pdf}, but for the PDFs of $\Sigma_{\star}$ and the fitted scale height of the {\sl tracer} population of stars, $h_z$, illustrated for the case of the metal-rich sub-population.}
 \label{fig:sh_pdf}
 \end{figure}

\begin{figure}
 \centering
 \includegraphics[width=0.9\textwidth]{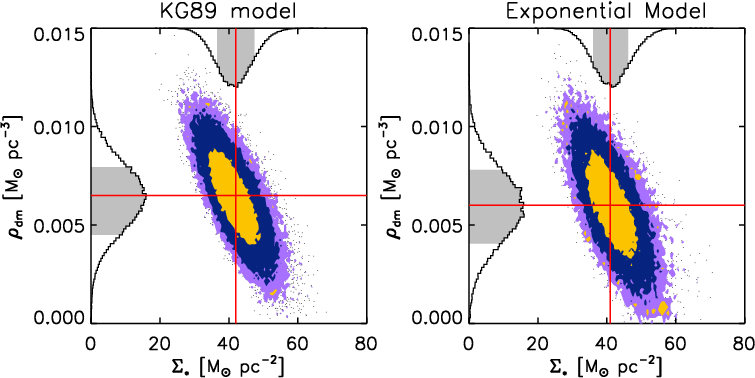}
 \caption{Analogous Figure to Fig.~\ref{fig:para_pdf}, but for the PDFs of $\rho_{\rm DM}$ and $\Sigma_{\star}$.}
 \label{fig:dm_pdf}
 \end{figure}

Fig.~\ref{fig:kz_force} shows $K_z(z)$ implied by our fits using the KG89b and exponential families for the `vertical force law'. The dashed fat lines show $K_z(z)$ for the most likely parameters, and the band of grey lines show a $1\sigma$ sampling of the PDF for $K_z(z)$. \textbf{Both of the KG89b and} the exponential families of $K_z(z)$ show a number of generic features in Fig.~\ref{fig:kz_force}: they start out with a finite value for small $|z|$, as we have fixed a prior contribution from a thin layer of the cold gas with $\sim 13~M_{\odot}\,{\rm pc}^{-2}$; then $K_z$ rises steeply to $\sim 300$~pc (reflecting the mass scale height of the stellar disk) and then flattens out; beyond $\sim 500$~pc the slower rise in $K_z$ reflects the dark matter halo term.

One the basis of $K_z(z)$, we can derive the total surface density at $|z| = 1.0$~kpc. We get
$\Sigma_{{\rm tot},|z|<1.0~{\rm kpc}}^{\rm KG89} = 67 \pm 6~M_{\odot}\,{\rm pc^{-2}}$ for the KG89 model and $\Sigma_{{\rm tot},|z|<1.0~{\rm kpc}}^{\rm Exp} = 66 \pm 8~M_{\odot}\,{\rm pc^{-2}}$ for the exponential model. It is clear that $\Sigma_{{\rm tot},|z|<1.0~{\rm kpc}}$ is robustly constrained, irrespective of the $K_z$ model.
Our modelling also constrains the mass scale height, $245^{+188}_{-245}$~pc for KG89 and $200^{+100}_{-200}$ for the exponential model (see Fig.~\ref{fig:para_pdf}). Within these uncertainties, this is consistent with the 180~pc from \citet{hill79} by analyzing A and F dwarfs and $390^{+330}_{-120}$~pc of \citet{sieb03}, found by using high resolution spectral data of red clump stars. Taken together this forms up the picture that there is a dominant mass layer near the disk mid-plane (presumably baryonic) that is rather flat.

For comparison with literature studies, we also explicitly estimate the total surface density of baryonic matter within $|z| < 1.1~{\rm kpc}$, $\Sigma_{{\rm baryonic},|z|<1.1~{\rm kpc}}=\Sigma_{\star}+\Sigma_{\rm gas}$, which we find to be
$\Sigma_{{\rm baryonic},|z|<1.1~{\rm kpc}}^{\rm KG89} = 55 \pm 5~M_{\odot}\,{\rm pc^{-2}}$, and $\Sigma_{{\rm baryonic},|z|<1.1~{\rm kpc}}^{\rm Exp} = 54 \pm 8~M_{\odot}\,{\rm pc^{-2}}$, respectively.
These values are slightly higher than $48 \pm 8~M_{\odot}\,{\rm pc}^{-2}$ derived by KG89b, but are in perfect agreement with the value of $52 \pm 13~M_{\odot}\,{\rm pc}^{-2}$ of FF94 and $53 \pm 6~M_{\odot}\,{\rm pc}^{-2}$ of HF04. 

The local mass densities of dark matter recovered by our models are $\rho_{\rm DM}^{\rm KG89} = 0.0065\pm0.0023~M_{\odot}\,{\rm pc^{-3}}$ ($0.25\pm0.09~{\rm GeV\,cm^{-3}}~$)\footnote{$1~{\rm GeV\,cm^{-3}} \simeq 0.0263158~M_{\odot}\,{\rm pc^{-3}}$} and $\rho_{\rm DM}^{\rm Exp} = 0.0060\pm0.0020~M_{\odot}\,{\rm pc^{-3}}$ ($0.23\pm0.08~{\rm GeV\,cm^{-3}}~$), marginalized over the other parameters..
We explored the parameter degeneracies by MCMC (see Fig.~\ref{fig:dm_pdf}), which recover the known degeneracy between surface mass density of stars and volume mass density of dark matter: lower $\Sigma_{\star}$ correlates with higher $\rho_{\rm DM}$, at an approximately constant total surface mass density.

\section{Discussion}
\label{sec:discussion}

We have presented an analysis of the vertical Galactic potential at the Solar
radius, drawing on $\sim$ 9000 K-dwarfs from SDSS/SEGUE. In many ways, the
analysis followed the Jeans equation approach initially laid out by KG89 and implemented by
several other groups in the mean-time \citep[][HF04, G12, BT12]{sieb03}. In comparison to most previous studies, our analysis has a number of new elements: we have a substantially larger sample than previous analyses; we have taken explicit account of the abundance-dependent spatial selection function of our sample for the analysis; we have simultaneously fit for several abundance-selected nearly-isothermal sub-populations that `feel' the same gravitational potential, and we have matched the kinematics and the spatial distribution simultaneously; we have explored to which
extent the choice of the functional form for $K_z$ affects the results.

As laid out in the previous Section, we find good constraints on
$\Sigma_{{\rm tot},|z|<1.0~{\rm kpc}}$ and some constrains on thickness of the disk {\it mass} layer: $z_h$:
$\Sigma_{{\rm tot},|z|<1.0~{\rm kpc}}=67 \pm 6~M_{\odot}\,{\rm pc^{-2}}$ and $z_h \lesssim 300$ pc, irrespective of the
functional form we assume for K$_z$. This is also among the first studies to find significant ($ > 2\sigma$) constraints on the local dark matter density, $\rho_{\rm DM} = 0.0065\pm0.0023~M_{\odot}\,{\rm pc^{-3}}$ ($0.25\pm0.09~{\rm GeV\,cm^{-3}}~$). Fig.~\ref{fig:dm_pdf} shows the expected degeneracy
between total surface density of visible matter and dark matter density: lower $\Sigma_{\star}$ corresponds to a higher $\rho_{\rm DM}$ for a given total surface density. However, these uncertainties do not include the systematic errors from misestimated photometric distances. For example, unrecognized binary contamination would make the inferred distances 10\% larger, and the distance estimation of \citet{an09} which we adopted in the present work is $\sim$ 9\% smaller than the one of \citet{ive08}. We explore distance systematics in particular for their implications on the dark matter density: if we systematically change all distances in the input data catalog by a $\pm$10\% and then go through the analysis steps of Sec.~3 is considered, the inferred change of $\rho_{\rm DM}$ is only $\pm$10\%.

\begin{figure}
 \centering
 \includegraphics[width=0.85\textwidth]{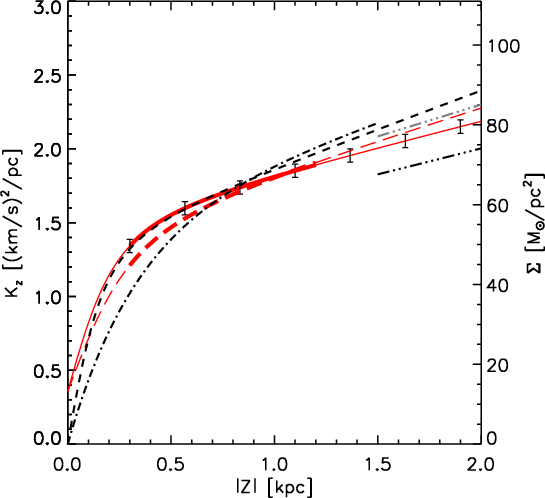}
 \caption{Comparison of the vertical force $K_z(z)$ and corresponding surface mass density $\Sigma_{{\rm tot},<|z|}$ implied by the best fits of the various model families to our data. The dark solid, and the long dashed red lines represent three cases of our predictions based on KG89 model family: the one with best fitting DM and the one with $\rho_{\rm DM} =0.008 M_{\odot}\,{\rm pc^{-3}}$ (from BT12), respectively. The 68\% uncertainty intervals on the surface-mass density are shown at a few representative points. The $|z|$ range where the lines are drawn thicker represent the $|z|$ range of our sample stars. Dashed line is the same model but from the prediction of KG89.}
 \label{fig:kz_compare}
 \end{figure}

Overall, our results on $K_z$ are consistent with earlier studies, as
Fig.~\ref{fig:kz_compare} shows, which compares the inferred total surface density $\Sigma_{\rm tot}(z)$ with previous studies.
The value derived by KG91 is $\Sigma_{{\rm tot},|z|<1.1~{\rm kpc}} = 71 \pm 6 ~M_{\odot}\,{\rm
pc^{-2}}$ and $\rho_{\rm DM} = 0.01 \pm 0.005
~M_{\odot}\,{\rm pc^{-3}}$.
In the calculation of HF04, $\Sigma_{{\rm tot},|z|<1.1~{\rm kpc}} = 74 \pm 6~M_{\odot}\,{\rm
pc^{-2}}$ and $\rho_{\rm DM} = 0.007~M_{\odot}\,{\rm pc^{-3}}$, whereas the latter results are in a good agreement with ours.

\begin{figure}
 \centering
 \includegraphics[width=0.85\textwidth]{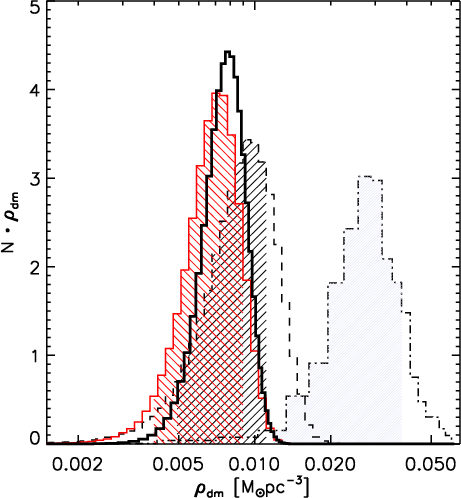}
 \caption{Comparison of normalized PDFs of $\rho_{\rm DM}$ derived by three independent works. The dashed dot line and gray shade represent the results of G12, dashed curve and shade are from the calculation of BT12, and red curve and shade are the results of our present work. Thicker black line means the joint PDF emerging from these three histogram.}
 \label{fig:rho_compare}
 \end{figure}

Fig.~\ref{fig:rho_compare} compares our inference for $\rho_{\rm DM}$ with two recent studies, while it agrees very well with BT12, our value is considerably lower than the $\rho_{\rm DM}$ inferred by G12, who found a value of $\rho_{\rm DM} = 0.025^{+0.014}_{-0.013}\,M_\odot\,{\rm pc}^{-2}$ ($0.85^{+0.57}_{-0.49}~{\rm GeV\,cm^{-3}}$), a considerably higher value than the present work (hereafter Z12), but  the two set of works are statistically consistent at  90\% confidence. G12 use a five times smaller sample over a comparable value; the sample has a simple selection function, but is not separated in abundance sub-bins. This may be the main reason to cause the difference in dynamic mass density of dark matter. In principle, the Jeans equation is linear, which means that the dynamical inferences should  not change whether the tracer population are mixing or not. However, splitting stellar populations according to their abundances can provide  stronger constraints on scale heights of vertical profile of tracers and hence place a stronger constraint on dark matter density. We use exponential $K_z$ model (Eq.~\ref{eq:kz_exp}) as an example to make a test, and find that the errors on $\rho_{\rm DM}$ are larger.

On the other hand, our estimates are in very good agreement (Fig.~\ref{fig:rho_compare}) with the recent determination of $\rho_{\rm DM}$ by BT12, who re-analyzed the data by \citet{moni12}. They used the assumption of $\partial V_c /\partial R =0$ to correct the model of \citet{moni12}, and derived a dark matter density of $\rho_{\rm DM} = 0.008\pm0.003~M_{\odot}\,{\rm pc^{-3}}$ ($0.30\pm0.11~{\rm GeV\,cm^{-3}}$).

Fig.~\ref{fig:rho_compare} puts the recent local $\rho_{\rm DM}$ estimate from $K_z$ constraints into context. The histogram with the gray shading in its center represents the results from G12, the one with black is from BT12 and the one with red is from Z12. The BT12 histogram reflects their $\rho_{\rm DM}$ PDF including systematic errors; for Z12 several of these error sources have been addressed systematically, and the distance uncertainties have been incorporated as a systematic error term in the Z12 histogram in Fig.~\ref{fig:rho_compare}. Therefore, the results shown in Fig.~\ref{fig:rho_compare} from Z12 and BT12 should be on comparable footing.
The joint PDF emerging from these three recent experiments is shown as the thick black histogram in Fig~\ref{fig:rho_compare}, indicating $\rho_{\rm DM}=0.0075 \pm 0.0021 ~M_{\odot}\,{\rm pc^{-3}}$ ($0.28\pm0.08~{\rm GeV\,cm^{-3}}~$).

One can put these `local' estimates of the dark matter density into the context of the expectations from the global fits of the Galactic dark matter halo: with  a spherical NFW cold dark matter density profile \citep{nav96} and
the parameters of \citet{xue08} (virial mass $M_{\rm vir} = 0.91^{+0.27}_{-0.18} \times 10^{12}~M_{\odot}$, virial radius $r_{\rm vir} = 267^{+24}_{-19}~{\rm kpc}$, and $c=12.0$), one would expect that the dark
matter density at $R_{\odot}$ is $\rho_{\rm DM}(R_{\odot}) = 0.0063~M_{\odot}\,{\rm pc^{-3}}$, which is in very good, perhaps even fortuitous,  agreement with our present work.

Taken together with the existing work, our results continue to point towards a picture that the local Galactic potential is dominated by a fairly thin layer of stars and gas plus a dark matter density that is in accord with global fits to Galactic halo. Specifically, we find no evidence for any significant amount of disk dark matter: our combined Z12+B12 estimate rules out the best-fit value of G12 of $\rho_\mathrm{DM} = 0.025\,\mathrm{M}_\odot\,\mathrm{pc}^{-3}$ at ~8$\sigma$ and our dynamical measurement of the surface density of baryonic matter of $\Sigma_\mathrm{baryonic} = 55\pm6\,\mathrm{M}_\odot\,\mathrm{pc}^{-2}$ is in good agreement with direct estimates through star counts and mapping of the local ISM. But note that due to the error bar of G12 our results are formally consistent with G12; therefore, whether a disk-like dark matter distribution at a normalization lower than that advocate by G12 cannot be ruled out, and should be tested further, as it contributes a qualitatively different scenario from an only-round halo.

It may seem surprising, why our limits we derive are not much tighter than those obtained by KG89b and subsequent work, based on smaller samples. This is in good part due to the fact that we used far fewer prior constraints on the models. E.g. we {\it fit} for the mass scale-height of the disk ($z_h$) rather than assume it; similarly, we fit for the scale height of the tracers simultaneously with the kinematic fit; we fit for dark matter density rather than assume it as a prior; and we have explored a range of functionl forms for $K_z$. Marginalizing over
these factors, apparently leads to a seemingly similar error on the parameters than other more restricted analysis with smaller samples.

\section{Conclusion}
\label{sec:conclusion}
We have analyzed the K-dwarfs from SDSS/SEGUE for determining the total surface mass density and dark matter density in the local Galactic disc. At first, we divide our sample into three sub-populations through their [Fe/H] and [$\alpha$/Fe]. After considering the spatial selection function, the Galactic vertical number density profiles for different sub-populations are inferred from star counting. Secondly, we use maximum likelihood and Gauss-Hermite series to calculate the vertical velocity dispersion profile of each sub-population. Then different parameterized $K_z$ forms are used to solve the `vertical' Jeans equation. We fit the observed vertical number density and vertical velocity dispersion profiles of the three sub-populations simultaneously, using MCMC to recover the PDFs of the parameters of the `vertical force law'.
In our results, for each sub-population, the vertical number density is approximately single exponential profile and the vertical velocity dispersion is nearly isothermal. The scale height of number density profile and the vertical velocity dispersion increase with the decline of metallicity. Presuming that there is a thin gas layer with $\Sigma_{\rm gas} = 13~M_{\odot}\,{\rm pc^{-2}}$, we derive a total surface mass density of $67 \pm 6~M_{\odot}\,{\rm pc^{-2}}$ at 1.0 kpc from the mid-plane, of which the contribution of all stars is $42 \pm 5 M_{\odot}\,{\rm pc^{-2}}$ and we infer a local dark matter density of $\rho_{\rm DM} = 0.0065 \pm 0.0023~M_{\odot}/{\rm pc^{-3}}$ ($0.25\pm0.09~{\rm GeV\,cm^{-3}}~$).

\acknowledgments
We would like to thank the anonymous referee for helpful comments, Constance Rockosi for the assistance of distance calculation, and Justin Read for discussions and suggestions. LZ acknowledges support of NSFC grant 10903012 and 11103034 and from the MPG-CAS student program. HWR, GvdV, and JB acknowledge partial support from Sonderforschungsbereich SFB 881 ``The Milky Way System" (subproject A3 and A7) funded by the German Research Foundation. JB was supported by NASA through Hubble Fellowship grant HST-HF-51285.01 from the Space Telescope Science Institute, which is operated by the Association of Universities for Research in Astronomy, Incorporated, under NASA contract NAS5-26555.

Funding for SDSS-III has been provided by the Alfred P. Sloan Foundation, the Participating Institutions, the National Science Foundation, and the U.S. Department of Energy Office of Science. The SDSS-III web site is http://www.sdss3.org/.

SDSS-III is managed by the Astrophysical Research Consortium for the Participating Institutions of the SDSS-III Collaboration including the University of Arizona, the Brazilian Participation Group, Brookhaven National Laboratory, University of Cambridge, Carnegie Mellon University, University of Florida, the French Participation Group, the German Participation Group, Harvard University, the Instituto de Astrofisica de Canarias, the Michigan State/Notre Dame/JINA Participation Group, Johns Hopkins University, Lawrence Berkeley National Laboratory, Max Planck Institute for Astrophysics, Max Planck Institute for Extraterrestrial Physics, New Mexico State University, New York University, Ohio State University, Pennsylvania State University, University of Portsmouth, Princeton University, the Spanish Participation Group, University of Tokyo, University of Utah, Vanderbilt University, University of Virginia, University of Washington, and Yale University.

\appendix

\section{Other Model Functional Forms for $K_z$}
Besides the two models for $K_z$ described in \S~\ref{subsec:kz_assum}, we also explore other three models to determine the total surface mass density and the mass density of dark matter, with the purpose of checking whether the astrophysical inferences depend on the choice of these functional forms:
\begin{enumerate}
  \item[iii)] {\bf Error function model}
       \begin{equation}
       \label{eq:kz_erf}
             K_z\,=\,2\,\pi\,G\left[\Sigma_{\star}\,{\rm erf}\left(\frac{z}{z_h}\right) +\Sigma_{\rm gas} \right]+ \,4\,\pi\,G\,\rho_{\rm DM}\,z
       \end{equation}
  \item[iv)] {\bf General model I}
        This form is an extension of KG model. We assume $K_z$ as the form of
        \begin{equation}
        \label{eq:kz_norm1}
              K_z\,=\,2\,\pi\,G\left[\Sigma_{\star}\left(\frac{z^\beta}{z^\gamma+z_h^\beta}\right)^
              {\frac{1}{\beta}}+\Sigma_{\rm gas}\right]+\,4\,\pi\,G\,\rho_{\rm DM}\,z
        \end{equation}
  \item[v)] {\bf General Model II}
        To reduce the number of free parameters, we set $\beta = \gamma$ and get
        \begin{equation}
        \label{eq:kz_norm2}
              K_z\,=\,2\,\pi\,G\left[\Sigma_{\star}\,\left(\frac{z^\beta}{z^\beta+z_h^\beta}\right)^
              {\frac{1}{\beta}}+\Sigma_{\rm gas}\right]+\,4\,\pi\,G\,\rho_{\rm DM}\,z
        \end{equation}
\end{enumerate}

We followed the same parameter estimation approach for these models and calculated the total $\chi^2$ between observed $\sigma_{z,{\rm obs}}$ and $\nu_{\star}$ and the model predictions. The error function model and the general model II yield similar results for the surface density of visible matter and dark matter density as the KG89 and exponetial models discussed in the main text. Only the general models I, where the exponents $\gamma$ and $\beta$ are independent yields to a degree of parameter degeneracy that makes the resulting $K_z(z)$ difficult to interpret.
But overall this confirms that among physically plausible families of $K_z$, the particular choice matters little.

\bibliographystyle{apj}
\bibliography{ref}

\begin{thebibliography}{46}
\expandafter\ifx\csname natexlab\endcsname\relax\def\natexlab#1{#1}\fi

\bibitem[{{Abazajian} {et~al.}(2009){Abazajian}, {Adelman-McCarthy},
  {Ag{\"u}eros}, {Allam}, {Allende Prieto}, {An}, {Anderson}, {Anderson},
  {Annis}, {Bahcall}, \& et~al.}]{aba09}
{Abazajian}, K.~N., {Adelman-McCarthy}, J.~K., {Ag{\"u}eros}, M.~A., {et~al.}
  2009, \apjs, 182, 543

\bibitem[{{Aihara} {et~al.}(2011{\natexlab{a}}){Aihara}, {Allende Prieto},
  {An}, {Anderson}, {Aubourg}, {Balbinot}, {Beers}, {Berlind}, {Bickerton},
  {Bizyaev}, {Blanton}, {Bochanski}, {Bolton}, {Bovy}, {Brandt}, {Brinkmann},
  {Brown}, {Brownstein}, {Busca}, {Campbell}, {Carr}, {Chen}, {Chiappini},
  {Comparat}, {Connolly}, {Cortes}, {Croft}, {Cuesta}, {da Costa}, {Davenport},
  {Dawson}, {Dhital}, {Ealet}, {Ebelke}, {Edmondson}, {Eisenstein},
  {Escoffier}, {Esposito}, {Evans}, {Fan}, {Femen{\'{\i}}a Castell{\'a}},
  {Font-Ribera}, {Frinchaboy}, {Ge}, {Gillespie}, {Gilmore}, {Gonz{\'a}lez
  Hern{\'a}ndez}, {Gott}, {Gould}, {Grebel}, {Gunn}, {Hamilton}, {Harding},
  {Harris}, {Hawley}, {Hearty}, {Ho}, {Hogg}, {Holtzman}, {Honscheid}, {Inada},
  {Ivans}, {Jiang}, {Johnson}, {Jordan}, {Jordan}, {Kazin}, {Kirkby}, {Klaene},
  {Knapp}, {Kneib}, {Kochanek}, {Koesterke}, {Kollmeier}, {Kron}, {Lampeitl},
  {Lang}, {Le Goff}, {Lee}, {Lin}, {Long}, {Loomis}, {Lucatello}, {Lundgren},
  {Lupton}, {Ma}, {MacDonald}, {Mahadevan}, {Maia}, {Makler}, {Malanushenko},
  {Malanushenko}, {Mandelbaum}, {Maraston}, {Margala}, {Masters}, {McBride},
  {McGehee}, {McGreer}, {M{\'e}nard}, {Miralda-Escud{\'e}}, {Morrison},
  {Mullally}, {Muna}, {Munn}, {Murayama}, {Myers}, {Naugle}, {Fausti Neto},
  {Cuong Nguyen}, {Nichol}, {O'Connell}, {Ogando}, {Olmstead}, {Oravetz},
  {Padmanabhan}, {Palanque-Delabrouille}, {Pan}, {Pandey}, {P{\^a}ris},
  {Percival}, {Petitjean}, {Pfaffenberger}, {Pforr}, {Phleps}, {Pichon},
  {Pieri}, {Prada}, {Price-Whelan}, {Raddick}, {Ramos}, {Reyl{\'e}}, {Rich},
  {Richards}, {Rix}, {Robin}, {Rocha-Pinto}, {Rockosi}, {Roe}, {Rollinde},
  {Ross}, {Ross}, {Rossetto}, {S{\'a}nchez}, {Sayres}, {Schlegel},
  {Schlesinger}, {Schmidt}, {Schneider}, {Sheldon}, {Shu}, {Simmerer},
  {Simmons}, {Sivarani}, {Snedden}, {Sobeck}, {Steinmetz}, {Strauss}, {Szalay},
  {Tanaka}, {Thakar}, {Thomas}, {Tinker}, {Tofflemire}, {Tojeiro}, {Tremonti},
  {Vandenberg}, {Vargas Maga{\~n}a}, {Verde}, {Vogt}, {Wake}, {Wang}, {Weaver},
  {Weinberg}, {White}, {White}, {Yanny}, {Yasuda}, {Yeche}, \&
  {Zehavi}}]{aiha11a}
{Aihara}, H., {Allende Prieto}, C., {An}, D., {et~al.} 2011{\natexlab{a}},
  \apjs, 193, 29

\bibitem[{{Aihara} {et~al.}(2011{\natexlab{b}}){Aihara}, {Allende Prieto},
  {An}, {Anderson}, {Aubourg}, {Balbinot}, {Beers}, {Berlind}, {Bickerton},
  {Bizyaev}, {Blanton}, {Bochanski}, {Bolton}, {Bovy}, {Brandt}, {Brinkmann},
  {Brown}, {Brownstein}, {Busca}, {Campbell}, {Carr}, {Chen}, {Chiappini},
  {Comparat}, {Connolly}, {Cortes}, {Croft}, {Cuesta}, {da Costa}, {Davenport},
  {Dawson}, {Dhital}, {Ealet}, {Ebelke}, {Edmondson}, {Eisenstein},
  {Escoffier}, {Esposito}, {Evans}, {Fan}, {Femen{\'{\i}}a Castell{\'a}},
  {Font-Ribera}, {Frinchaboy}, {Ge}, {Gillespie}, {Gilmore}, {Gonz{\'a}lez
  Hern{\'a}ndez}, {Gott}, {Gould}, {Grebel}, {Gunn}, {Hamilton}, {Harding},
  {Harris}, {Hawley}, {Hearty}, {Ho}, {Hogg}, {Holtzman}, {Honscheid}, {Inada},
  {Ivans}, {Jiang}, {Johnson}, {Jordan}, {Jordan}, {Kazin}, {Kirkby}, {Klaene},
  {Knapp}, {Kneib}, {Kochanek}, {Koesterke}, {Kollmeier}, {Kron}, {Lampeitl},
  {Lang}, {Le Goff}, {Lee}, {Lin}, {Long}, {Loomis}, {Lucatello}, {Lundgren},
  {Lupton}, {Ma}, {MacDonald}, {Mahadevan}, {Maia}, {Makler}, {Malanushenko},
  {Malanushenko}, {Mandelbaum}, {Maraston}, {Margala}, {Masters}, {McBride},
  {McGehee}, {McGreer}, {M{\'e}nard}, {Miralda-Escud{\'e}}, {Morrison},
  {Mullally}, {Muna}, {Munn}, {Murayama}, {Myers}, {Naugle}, {Fausti Neto},
  {Cuong Nguyen}, {Nichol}, {O'Connell}, {Ogando}, {Olmstead}, {Oravetz},
  {Padmanabhan}, {Palanque-Delabrouille}, {Pan}, {Pandey}, {P{\^a}ris},
  {Percival}, {Petitjean}, {Pfaffenberger}, {Pforr}, {Phleps}, {Pichon},
  {Pieri}, {Prada}, {Price-Whelan}, {Raddick}, {Ramos}, {Reyl{\'e}}, {Rich},
  {Richards}, {Rix}, {Robin}, {Rocha-Pinto}, {Rockosi}, {Roe}, {Rollinde},
  {Ross}, {Ross}, {Rossetto}, {S{\'a}nchez}, {Sayres}, {Schlegel},
  {Schlesinger}, {Schmidt}, {Schneider}, {Sheldon}, {Shu}, {Simmerer},
  {Simmons}, {Sivarani}, {Snedden}, {Sobeck}, {Steinmetz}, {Strauss}, {Szalay},
  {Tanaka}, {Thakar}, {Thomas}, {Tinker}, {Tofflemire}, {Tojeiro}, {Tremonti},
  {Vandenberg}, {Vargas Maga{\~n}a}, {Verde}, {Vogt}, {Wake}, {Wang}, {Weaver},
  {Weinberg}, {White}, {White}, {Yanny}, {Yasuda}, {Yeche}, \&
  {Zehavi}}]{aiha11b}
{Aihara}, H., {Allende Prieto}, C., {An}, D., {et~al.} 2011{\natexlab{b}},
  \apjs, 195, 26

\bibitem[{{Allende Prieto} {et~al.}(2008){Allende Prieto}, {Sivarani}, {Beers},
  {Lee}, {Koesterke}, {Shetrone}, {Sneden}, {Lambert}, {Wilhelm}, {Rockosi},
  {Lai}, {Yanny}, {Ivans}, {Johnson}, {Aoki}, {Bailer-Jones}, \& {Re
  Fiorentin}}]{all08}
{Allende Prieto}, C., {Sivarani}, T., {Beers}, T.~C., {et~al.} 2008, \aj, 136,
  2070

\bibitem[{{An} {et~al.}(2009){An}, {Pinsonneault}, {Masseron}, {Delahaye},
  {Johnson}, {Terndrup}, {Beers}, {Ivans}, \& {Ivezi{\'c}}}]{an09}
{An}, D., {Pinsonneault}, M.~H., {Masseron}, T., {et~al.} 2009, \apj, 700, 523

\bibitem[{{Bienaym{\'e}} {et~al.}(2006){Bienaym{\'e}}, {Soubiran}, {Mishenina},
  {Kovtyukh}, \& {Siebert}}]{bien06}
{Bienaym{\'e}}, O., {Soubiran}, C., {Mishenina}, T.~V., {Kovtyukh}, V.~V., \&
  {Siebert}, A. 2006, \aap, 446, 933

\bibitem[{{Binney} \& {Tremaine}(2008)}]{bin08}
{Binney}, J. \& {Tremaine}, S. 2008, {Galactic Dynamics: Second Edition}, ed.
  {Binney, J.~\& Tremaine, S.} (Princeton University Press)

\bibitem[{{Bovy} {et~al.}(2012{\natexlab{a}}){Bovy}, {Allende Prieto}, {Beers},
  {Bizyaev}, {da Costa}, {Cunha}, {Ebelke}, {Eisenstein}, {Frinchaboy}, {Elia
  Garc{\'{\i}}a P{\'e}rez}, {Girardi}, {Hearty}, {Hogg}, {Holtzman}, {Maia},
  {Majewski}, {Malanushenko}, {Malanushenko}, {M{\'e}sz{\'a}ros}, {Nidever},
  {O'Connell}, {O'Donnell}, {Oravetz}, {Pan}, {Rocha-Pinto}, {Schiavon},
  {Schneider}, {Schultheis}, {Skrutskie}, {Smith}, {Weinberg}, {Wilson}, \&
  {Zasowski}}]{bovy12e}
{Bovy}, J., {Allende Prieto}, C., {Beers}, T.~C., {et~al.} 2012{\natexlab{a}},
  ArXiv e-prints

\bibitem[{{Bovy} {et~al.}(2012{\natexlab{b}}){Bovy}, {Rix}, \&
  {Hogg}}]{bovy12a}
{Bovy}, J., {Rix}, H.-W., \& {Hogg}, D.~W. 2012{\natexlab{b}}, \apj, 751, 131

\bibitem[{{Bovy} {et~al.}(2012{\natexlab{c}}){Bovy}, {Rix}, {Hogg}, {Beers},
  {Lee}, \& {Zhang}}]{bovy12b}
{Bovy}, J., {Rix}, H.-W., {Hogg}, D.~W., {et~al.} 2012{\natexlab{c}}, \apj, 755

\bibitem[{{Bovy} {et~al.}(2012{\natexlab{d}}){Bovy}, {Rix}, {Liu}, {Hogg},
  {Beers}, \& {Lee}}]{bovy12c}
{Bovy}, J., {Rix}, H.-W., {Liu}, C., {et~al.} 2012{\natexlab{d}}, \apj, 753,
  148

\bibitem[{{Bovy} \& {Tremaine}(2012)}]{bovy12d}
{Bovy}, J. \& {Tremaine}, S. 2012, \apj, 756, 89

\bibitem[{{Dehnen} \& {Binney}(1998)}]{dehn98}
{Dehnen}, W. \& {Binney}, J.~J. 1998, \mnras, 298, 387

\bibitem[{{Flynn} \& {Fuchs}(1994)}]{flyn94}
{Flynn}, C. \& {Fuchs}, B. 1994, \mnras, 270, 471

\bibitem[{{Flynn} {et~al.}(2006){Flynn}, {Holmberg}, {Portinari}, {Fuchs}, \&
  {Jahrei{\ss}}}]{flyn06}
{Flynn}, C., {Holmberg}, J., {Portinari}, L., {Fuchs}, B., \& {Jahrei{\ss}}, H.
  2006, \mnras, 372, 1149

\bibitem[{{Garbari} {et~al.}(2012){Garbari}, {Liu}, {Read}, \& {Lake}}]{garb12}
{Garbari}, S., {Liu}, C., {Read}, J.~I., \& {Lake}, G. 2012, \mnras, 425, 1445

\bibitem[{{Garbari} {et~al.}(2011){Garbari}, {Read}, \& {Lake}}]{gar11}
{Garbari}, S., {Read}, J.~I., \& {Lake}, G. 2011, \mnras, 416, 2318

\bibitem[{{Gerhard}(1993)}]{ger93}
{Gerhard}, O.~E. 1993, \mnras, 265, 213

\bibitem[{{Hill} {et~al.}(1979){Hill}, {Hilditch}, \& {Barnes}}]{hill79}
{Hill}, G., {Hilditch}, R.~W., \& {Barnes}, J.~V. 1979, \mnras, 186, 813

\bibitem[{{Holmberg} \& {Flynn}(2004)}]{holm04}
{Holmberg}, J. \& {Flynn}, C. 2004, \mnras, 352, 440

\bibitem[{{Ivezi{\'c}} {et~al.}(2008){Ivezi{\'c}}, {Sesar}, {Juri{\'c}},
  {Bond}, {Dalcanton}, {Rockosi}, {Yanny}, {Newberg}, {Beers}, {Allende
  Prieto}, {Wilhelm}, {Lee}, {Sivarani}, {Norris}, {Bailer-Jones}, {Re
  Fiorentin}, {Schlegel}, {Uomoto}, {Lupton}, {Knapp}, {Gunn}, {Covey},
  {Smith}, {Miknaitis}, {Doi}, {Tanaka}, {Fukugita}, {Kent}, {Finkbeiner},
  {Munn}, {Pier}, {Quinn}, {Hawley}, {Anderson}, {Kiuchi}, {Chen}, {Bushong},
  {Sohi}, {Haggard}, {Kimball}, {Barentine}, {Brewington}, {Harvanek},
  {Kleinman}, {Krzesinski}, {Long}, {Nitta}, {Snedden}, {Lee}, {Harris},
  {Brinkmann}, {Schneider}, \& {York}}]{ive08}
{Ivezi{\'c}}, {\v Z}., {Sesar}, B., {Juri{\'c}}, M., {et~al.} 2008, \apj, 684,
  287

\bibitem[{{Korchagin} {et~al.}(2003){Korchagin}, {Girard}, {Borkova},
  {Dinescu}, \& {van Altena}}]{korc03}
{Korchagin}, V.~I., {Girard}, T.~M., {Borkova}, T.~V., {Dinescu}, D.~I., \&
  {van Altena}, W.~F. 2003, \aj, 126, 2896

\bibitem[{{Kuijken} \& {Gilmore}(1989{\natexlab{a}})}]{kui89ii}
{Kuijken}, K. \& {Gilmore}, G. 1989{\natexlab{a}}, \mnras, 239, 605

\bibitem[{{Kuijken} \& {Gilmore}(1989{\natexlab{b}})}]{kui89iii}
{Kuijken}, K. \& {Gilmore}, G. 1989{\natexlab{b}}, \mnras, 239, 651

\bibitem[{{Kuijken} \& {Gilmore}(1989{\natexlab{c}})}]{kui89i}
{Kuijken}, K. \& {Gilmore}, G. 1989{\natexlab{c}}, \mnras, 239, 571

\bibitem[{{Kuijken} \& {Gilmore}(1991)}]{kuij91}
{Kuijken}, K. \& {Gilmore}, G. 1991, \apjl, 367, L9

\bibitem[{{Lee} {et~al.}(2011){Lee}, {Beers}, {An}, {Ivezi{\'c}}, {Just},
  {Rockosi}, {Morrison}, {Johnson}, {Sch{\"o}nrich}, {Bird}, {Yanny},
  {Harding}, \& {Rocha-Pinto}}]{lee11}
{Lee}, Y.~S., {Beers}, T.~C., {An}, D., {et~al.} 2011, \apj, 738, 187

\bibitem[{{Lee} {et~al.}(2008{\natexlab{a}}){Lee}, {Beers}, {Sivarani},
  {Allende Prieto}, {Koesterke}, {Wilhelm}, {Re Fiorentin}, {Bailer-Jones},
  {Norris}, {Rockosi}, {Yanny}, {Newberg}, {Covey}, {Zhang}, \& {Luo}}]{lee08a}
{Lee}, Y.~S., {Beers}, T.~C., {Sivarani}, T., {et~al.} 2008{\natexlab{a}}, \aj,
  136, 2022

\bibitem[{{Lee} {et~al.}(2008{\natexlab{b}}){Lee}, {Beers}, {Sivarani},
  {Johnson}, {An}, {Wilhelm}, {Allende Prieto}, {Koesterke}, {Re Fiorentin},
  {Bailer-Jones}, {Norris}, {Yanny}, {Rockosi}, {Newberg}, {Cudworth}, \&
  {Pan}}]{lee08b}
{Lee}, Y.~S., {Beers}, T.~C., {Sivarani}, T., {et~al.} 2008{\natexlab{b}}, \aj,
  136, 2050

\bibitem[{{Liu} \& {van de Ven}(2012)}]{liu12}
{Liu}, C. \& {van de Ven}, G. 2012, \mnras, in press

\bibitem[{{Moni Bidin} {et~al.}(2012){Moni Bidin}, {Carraro}, {M{\'e}ndez}, \&
  {Smith}}]{moni12}
{Moni Bidin}, C., {Carraro}, G., {M{\'e}ndez}, R.~A., \& {Smith}, R. 2012,
  \apj, 751, 30

\bibitem[{{Navarro} {et~al.}(1996){Navarro}, {Frenk}, \& {White}}]{nav96}
{Navarro}, J.~F., {Frenk}, C.~S., \& {White}, S.~D.~M. 1996, \apj, 462, 563

\bibitem[{{Oort}(1932)}]{oort32}
{Oort}, J.~H. 1932, \bain, 6, 249

\bibitem[{{Oort}(1960)}]{oort60}
{Oort}, J.~H. 1960, \bain, 15, 45

\bibitem[{{Press} {et~al.}(2007){Press}, {Teukolsky}, {Vetterling}, \&
  {Flannery}}]{pre07}
{Press}, W.~H., {Teukolsky}, S.~A., {Vetterling}, W.~T., \& {Flannery}, B.~P.
  2007, {Numerical Recipes. The art of scientific computing}, ed. {Press,
  W.~H., Teukolsky, S.~A., Vetterling, W.~T., \& Flannery, B.~P. }

\bibitem[{{Reid}(1993)}]{reid93}
{Reid}, M.~J. 1993, \araa, 31, 345

\bibitem[{{Schlegel} {et~al.}(1998){Schlegel}, {Finkbeiner}, \&
  {Davis}}]{schl98}
{Schlegel}, D.~J., {Finkbeiner}, D.~P., \& {Davis}, M. 1998, \apj, 500, 525

\bibitem[{{Siebert} {et~al.}(2003){Siebert}, {Bienaym{\'e}}, \&
  {Soubiran}}]{sieb03}
{Siebert}, A., {Bienaym{\'e}}, O., \& {Soubiran}, C. 2003, \aap, 399, 531

\bibitem[{{Smolinski} {et~al.}(2011){Smolinski}, {Lee}, {Beers}, {An},
  {Bickerton}, {Johnson}, {Loomis}, {Rockosi}, {Sivarani}, \& {Yanny}}]{smo11}
{Smolinski}, J.~P., {Lee}, Y.~S., {Beers}, T.~C., {et~al.} 2011, \aj, 141, 89

\bibitem[{{Stoughton} {et~al.}(2002){Stoughton}, {Lupton}, {Bernardi},
  {Blanton}, {Burles}, {Castander}, {Connolly}, {Eisenstein}, {Frieman},
  {Hennessy}, {Hindsley}, {Ivezi{\'c}}, {Kent}, {Kunszt}, {Lee}, {Meiksin},
  {Munn}, {Newberg}, {Nichol}, {Nicinski}, {Pier}, {Richards}, {Richmond},
  {Schlegel}, {Smith}, {Strauss}, {SubbaRao}, {Szalay}, {Thakar}, {Tucker},
  {Vanden Berk}, {Yanny}, {Adelman}, {Anderson}, {Anderson}, {Annis},
  {Bahcall}, {Bakken}, {Bartelmann}, {Bastian}, {Bauer}, {Berman},
  {B{\"o}hringer}, {Boroski}, {Bracker}, {Briegel}, {Briggs}, {Brinkmann},
  {Brunner}, {Carey}, {Carr}, {Chen}, {Christian}, {Colestock}, {Crocker},
  {Csabai}, {Czarapata}, {Dalcanton}, {Davidsen}, {Davis}, {Dehnen},
  {Dodelson}, {Doi}, {Dombeck}, {Donahue}, {Ellman}, {Elms}, {Evans}, {Eyer},
  {Fan}, {Federwitz}, {Friedman}, {Fukugita}, {Gal}, {Gillespie}, {Glazebrook},
  {Gray}, {Grebel}, {Greenawalt}, {Greene}, {Gunn}, {de Haas}, {Haiman},
  {Haldeman}, {Hall}, {Hamabe}, {Hansen}, {Harris}, {Harris}, {Harvanek},
  {Hawley}, {Hayes}, {Heckman}, {Helmi}, {Henden}, {Hogan}, {Hogg}, {Holmgren},
  {Holtzman}, {Huang}, {Hull}, {Ichikawa}, {Ichikawa}, {Johnston}, {Kauffmann},
  {Kim}, {Kimball}, {Kinney}, {Klaene}, {Kleinman}, {Klypin}, {Knapp},
  {Korienek}, {Krolik}, {Kron}, {Krzesi{\'n}ski}, {Lamb}, {Leger},
  {Limmongkol}, {Lindenmeyer}, {Long}, {Loomis}, {Loveday}, {MacKinnon},
  {Mannery}, {Mantsch}, {Margon}, {McGehee}, {McKay}, {McLean}, {Menou},
  {Merelli}, {Mo}, {Monet}, {Nakamura}, {Narayanan}, {Nash}, {Neilsen},
  {Newman}, {Nitta}, {Odenkirchen}, {Okada}, {Okamura}, {Ostriker}, {Owen},
  {Pauls}, {Peoples}, {Peterson}, {Petravick}, {Pope}, {Pordes}, {Postman},
  {Prosapio}, {Quinn}, {Rechenmacher}, {Rivetta}, {Rix}, {Rockosi}, {Rosner},
  {Ruthmansdorfer}, {Sandford}, {Schneider}, {Scranton}, {Sekiguchi}, {Sergey},
  {Sheth}, {Shimasaku}, {Smee}, {Snedden}, {Stebbins}, {Stubbs}, {Szapudi},
  {Szkody}, {Szokoly}, {Tabachnik}, {Tsvetanov}, {Uomoto}, {Vogeley}, {Voges},
  {Waddell}, {Walterbos}, {Wang}, {Watanabe}, {Weinberg}, {White}, {White},
  {Wilhite}, {Wolfe}, {Yasuda}, {York}, {Zehavi}, \& {Zheng}}]{stou02}
{Stoughton}, C., {Lupton}, R.~H., {Bernardi}, M., {et~al.} 2002, \aj, 123, 485

\bibitem[{{van de Ven} {et~al.}(2006){van de Ven}, {van den Bosch}, {Verolme},
  \& {de Zeeuw}}]{van06}
{van de Ven}, G., {van den Bosch}, R.~C.~E., {Verolme}, E.~K., \& {de Zeeuw},
  P.~T. 2006, \aap, 445, 513

\bibitem[{{van der Marel} \& {Franx}(1993)}]{van93}
{van der Marel}, R.~P. \& {Franx}, M. 1993, \apj, 407, 525

\bibitem[{{Weber} \& {de Boer}(2010)}]{webe10}
{Weber}, M. \& {de Boer}, W. 2010, \aap, 509, A25

\bibitem[{{Xue} {et~al.}(2008){Xue}, {Rix}, {Zhao}, {Re Fiorentin}, {Naab},
  {Steinmetz}, {van den Bosch}, {Beers}, {Lee}, {Bell}, {Rockosi}, {Yanny},
  {Newberg}, {Wilhelm}, {Kang}, {Smith}, \& {Schneider}}]{xue08}
{Xue}, X.~X., {Rix}, H.~W., {Zhao}, G., {et~al.} 2008, \apj, 684, 1143

\bibitem[{{Yanny} {et~al.}(2009){Yanny}, {Rockosi}, {Newberg}, {Knapp},
  {Adelman-McCarthy}, {Alcorn}, {Allam}, {Allende Prieto}, {An}, {Anderson},
  {Anderson}, {Bailer-Jones}, {Bastian}, {Beers}, {Bell}, {Belokurov},
  {Bizyaev}, {Blythe}, {Bochanski}, {Boroski}, {Brinchmann}, {Brinkmann},
  {Brewington}, {Carey}, {Cudworth}, {Evans}, {Evans}, {Gates}, {G{\"a}nsicke},
  {Gillespie}, {Gilmore}, {Nebot Gomez-Moran}, {Grebel}, {Greenwell}, {Gunn},
  {Jordan}, {Jordan}, {Harding}, {Harris}, {Hendry}, {Holder}, {Ivans},
  {Ivezi{\v c}}, {Jester}, {Johnson}, {Kent}, {Kleinman}, {Kniazev},
  {Krzesinski}, {Kron}, {Kuropatkin}, {Lebedeva}, {Lee}, {French Leger},
  {L{\'e}pine}, {Levine}, {Lin}, {Long}, {Loomis}, {Lupton}, {Malanushenko},
  {Malanushenko}, {Margon}, {Martinez-Delgado}, {McGehee}, {Monet}, {Morrison},
  {Munn}, {Neilsen}, {Nitta}, {Norris}, {Oravetz}, {Owen}, {Padmanabhan},
  {Pan}, {Peterson}, {Pier}, {Platson}, {Re Fiorentin}, {Richards}, {Rix},
  {Schlegel}, {Schneider}, {Schreiber}, {Schwope}, {Sibley}, {Simmons},
  {Snedden}, {Allyn Smith}, {Stark}, {Stauffer}, {Steinmetz}, {Stoughton},
  {SubbaRao}, {Szalay}, {Szkody}, {Thakar}, {Thirupathi}, {Tucker}, {Uomoto},
  {Vanden Berk}, {Vidrih}, {Wadadekar}, {Watters}, {Wilhelm}, {Wyse}, {Yarger},
  \& {Zucker}}]{yann09}
{Yanny}, B., {Rockosi}, C., {Newberg}, H.~J., {et~al.} 2009, \aj, 137, 4377

\bibitem[{{York} {et~al.}(2000){York}, {Adelman}, {Anderson}, {Anderson},
  {Annis}, {Bahcall}, {Bakken}, {Barkhouser}, {Bastian}, {Berman}, {Boroski},
  {Bracker}, {Briegel}, {Briggs}, {Brinkmann}, {Brunner}, {Burles}, {Carey},
  {Carr}, {Castander}, {Chen}, {Colestock}, {Connolly}, {Crocker}, {Csabai},
  {Czarapata}, {Davis}, {Doi}, {Dombeck}, {Eisenstein}, {Ellman}, {Elms},
  {Evans}, {Fan}, {Federwitz}, {Fiscelli}, {Friedman}, {Frieman}, {Fukugita},
  {Gillespie}, {Gunn}, {Gurbani}, {de Haas}, {Haldeman}, {Harris}, {Hayes},
  {Heckman}, {Hennessy}, {Hindsley}, {Holm}, {Holmgren}, {Huang}, {Hull},
  {Husby}, {Ichikawa}, {Ichikawa}, {Ivezi{\'c}}, {Kent}, {Kim}, {Kinney},
  {Klaene}, {Kleinman}, {Kleinman}, {Knapp}, {Korienek}, {Kron}, {Kunszt},
  {Lamb}, {Lee}, {Leger}, {Limmongkol}, {Lindenmeyer}, {Long}, {Loomis},
  {Loveday}, {Lucinio}, {Lupton}, {MacKinnon}, {Mannery}, {Mantsch}, {Margon},
  {McGehee}, {McKay}, {Meiksin}, {Merelli}, {Monet}, {Munn}, {Narayanan},
  {Nash}, {Neilsen}, {Neswold}, {Newberg}, {Nichol}, {Nicinski}, {Nonino},
  {Okada}, {Okamura}, {Ostriker}, {Owen}, {Pauls}, {Peoples}, {Peterson},
  {Petravick}, {Pier}, {Pope}, {Pordes}, {Prosapio}, {Rechenmacher}, {Quinn},
  {Richards}, {Richmond}, {Rivetta}, {Rockosi}, {Ruthmansdorfer}, {Sandford},
  {Schlegel}, {Schneider}, {Sekiguchi}, {Sergey}, {Shimasaku}, {Siegmund},
  {Smee}, {Smith}, {Snedden}, {Stone}, {Stoughton}, {Strauss}, {Stubbs},
  {SubbaRao}, {Szalay}, {Szapudi}, {Szokoly}, {Thakar}, {Tremonti}, {Tucker},
  {Uomoto}, {Vanden Berk}, {Vogeley}, {Waddell}, {Wang}, {Watanabe},
  {Weinberg}, {Yanny}, {Yasuda}, \& {SDSS Collaboration}}]{york00}
{York}, D.~G., {Adelman}, J., {Anderson}, Jr., J.~E., {et~al.} 2000, \aj, 120,
  1579

\end{thebibliography}

\clearpage


 \clearpage

\clearpage
 
 \clearpage

 \clearpage

 \clearpage

 \clearpage

 \clearpage

 \clearpage

 \clearpage

 \clearpage

 \clearpage

 \clearpage

\begin{table}
\begin{center}
\caption{Fitting results of the two models, with $\Sigma_{\rm gas} = 13~ M_{\odot}\,{\rm pc}^{-2}$ \citep{flyn06}}
\label{tab:result}
\begin{tabular}{lcc}
\tableline\tableline
                                                       & KG89 model (Eq.~\ref{eq:kz_kg89}) &  Exponential model (Eq.~\ref{eq:kz_exp})\\
\tableline
 $\Sigma_{\star}$ [$M_{\odot}\,{\rm pc}^{-2}$]  & 42 $\pm$ 6          & 41 $\pm$ 5          \\
 $\rho_{\rm DM}$ [$M_{\odot}\,{\rm pc}^{-3}$]          & 0.0064 $\pm$ 0.0023 & 0.0060 $\pm$ 0.0020 \\
  $z_h$ [pc]                                           & $245^{+188}_{-245}$ & $200^{+100}_{-200}$ \\
  $h_{z,1}$ \tablenotemark{a} [pc]                     & 259  $\pm$ 12       & 260  $\pm$ 15       \\
  $h_{z,2}$ \tablenotemark{a} [pc]                     & 450  $\pm$ 26       & 465  $\pm$ 33       \\
  $h_{z,3}$ \tablenotemark{a} [pc]                     & 852  $\pm$ 30       & 910  $\pm$ 71       \\
\tableline
  $\sigma_{0,1}$ \tablenotemark{a} [km/s]              & 15.4 $\pm$ 1.3      & 15.8 $\pm$ 1.3      \\
  $\sigma_{0,2}$ \tablenotemark{a} [km/s]              & 23.0 $\pm$ 2.0      & 23.6 $\pm$ 2.0      \\
  $\sigma_{0,3}$ \tablenotemark{a} [km/s]              & 34.2 $\pm$ 2.2      & 35.8 $\pm$ 2.2      \\
\tableline
  $\chi^2_{\sigma_z}$                                  & 25.23               & 26.36               \\
  $\chi^2_{\nu_{\star}}$                               & 15.80               & 14.54               \\
  $\chi^2_{\rm tot}$                                   & 41.03               & 40.90               \\
\tableline
\end{tabular}
\tablenotetext{a}{Number 1, 2, and 3 represent metal-rich, intermediate metallicity, and metal-poor sub-sample,respectively.}
\end{center}
\end{table}

\end{document}